\documentclass[prd,aps,preprintnumbers,floats,floatfix,superscriptaddress,preprintnumbers,
showpacs,eqsecnum,nofootinbib,twocolumn]{revtex4-1}
\usepackage[utf8]{inputenc}
\usepackage{latexsym,array,theorem,mathrsfs,bm,float}
\usepackage{psfrag}
\usepackage{amsfonts,amsmath,amssymb,latexsym,array,afterpage,
theorem,mathrsfs,bm,float,epsfig,color,graphicx,tabularx,here,multirow}

\newcommand{\nn}{\nonumber \\}
\newcommand{\bea}{\begin{eqnarray}}
\newcommand{\ena}{\end{eqnarray}}
\newcommand{\beann}{\begin{eqnarray*}}
\newcommand{\enann}{\end{eqnarray*}}

\newcommand{\gsim}{\, \mbox{\raisebox{-1.ex}
{$\stackrel{\textstyle>}{\textstyle\sim}$}}\,}
\newcommand{\lsim}{\, \mbox{\raisebox{-1.ex}
{$\stackrel{\textstyle<}{\textstyle\sim}$}}\,}
\newcommand{\vect}[1]{\!\!\!\mbox{\,~\boldmath $#1$}}

\newcommand{\GB}{\overset{\scriptscriptstyle (0)}{G}}

\newcommand{\gB}{\overset{\scriptscriptstyle  (0)}{g}}

\newcommand{\Rff}{\overset{\scriptscriptstyle (2)}{R}}

\newcommand{\nablaB}{\overset{\scriptscriptstyle  (0)}{\nabla}}

\begin{document}
\baselineskip=12pt

\preprint{WU-AP/1705/17} 

\title{Massive Graviton Geons
}
\author{Katsuki \sc{Aoki}}
\email{katsuki-a12@gravity.phys.waseda.ac.jp}
\affiliation{
Department of Physics, Waseda University,
Shinjuku, Tokyo 169-8555, Japan
}

\author{Kei-ichi \sc{Maeda}}
\email{maeda@waseda.ac.jp}
\affiliation{
Department of Physics, Waseda University,
Shinjuku, Tokyo 169-8555, Japan
}

\author{Yosuke \sc{Misonoh}}
\email{misonoh@aoni.waseda.jp}
\affiliation{
Department of Physics, Waseda University,
Shinjuku, Tokyo 169-8555, Japan
}

\author{Hirotada \sc{Okawa}}
\email{okawa@heap.phys.waseda.ac.jp}
\affiliation{Yukawa Institute for Theoretical Physics, Kyoto University, Kyoto 606-8502, Japan}
\affiliation{
Department of Physics, Waseda University,
Shinjuku, Tokyo 169-8555, Japan
}

\date{\today}

\begin{abstract}
We find vacuum solutions such that massive gravitons are confined in a local spacetime region by their gravitational energy in asymptotically flat spacetimes in the context of the bigravity theory. We call such self-gravitating objects massive graviton geons. The basic equations can be reduced to the Schr\"{o}dinger-Poisson equations with the tensor ``wavefunction'' in the Newtonian limit. We obtain a non-spherically symmetric solution with $j=2,\ell=0$ as well as a spherically symmetric solution with $j=0,\ell=2$ in this system where $j$ is the total angular momentum quantum number and $\ell$ is the orbital angular momentum quantum number, respectively. The energy eigenvalue of the Schr\"{o}dinger equation in the non-spherical solution is smaller than that in the spherical solution. We then study the perturbative stability of the spherical solution and find that there is an unstable mode in the quadrupole mode perturbations which may be interpreted as the transition mode to the non-spherical solution. The results suggest that the non-spherically symmetric solution is the ground state of the massive graviton geon. The massive graviton geons may decay in time due to emissions of gravitational waves but this timescale can be quite long when the massive gravitons are non-relativistic and then the geons can be long-lived. We also argue possible prospects of the massive graviton geons: applications to the ultralight dark matter scenario, nonlinear (in)stability of the Minkowski spacetime, and a quantum transition of the spacetime.
\end{abstract}



\maketitle

\section{Introduction}
Recent observations found gravitational waves from black hole mergers in which a few percent of the energy of the system is radiated by the gravitational waves \cite{Abbott:2016blz,TheLIGOScientific:2016pea,Abbott:2017vtc,Abbott:2017oio}. In this way, it is well-known that the gravitational waves have their energy and change the background geometry. Therefore, it seems possible that the gravitational waves are gravitationally bounded in a local region of the spacetime by their own gravitational energy. This time-dependent self-gravitating object is called a gravitational \emph{geon}, short for \emph{gravitational-electromagnetic entity}, which was introduced by Wheeler~\cite{Wheeler:1955zz}. Although a gravitational geon can be constructed in asymptotically flat spacetimes~\cite{Brill:1964zz,Anderson:1996pu}, the geon is not exactly periodic
in time \cite{Trautman:1957zz,Detweiler:1993eq,gibbons1984absence,Tod:2009em,Bicak:2010xp,Bicak:2010tt,Alexakis:2015ara,Enriquez:2016klf} and it has a finite lifetime. However, the instability of geons could be remedied in asymptotically anti-de Sitter (AdS) spacetimes since the AdS boundary can confine particles and waves as in a box. Gravitational geons in asymptotically AdS spacetimes are constructed in \cite{Dias:2012tq,Horowitz:2014hja,Martinon:2017uyo}.

In the present paper, we consider asymptotically flat spacetimes and construct a gravitational geon of massive gravitons.
The extra ingredient in the present analysis is a mass of a graviton. Although a graviton in general relativity is massless, theories with a massive graviton have received much attentions. For instance, models with extra-dimensions predict the existence of massive gravitons as well as a massless graviton as Kaluza-Klein modes. In other examples, the massive graviton has been discussed to give rise to the cosmic accelerating expansion  (see \cite{deRham:2014zqa,Schmidt-May:2015vnx} for reviews and references therein) and recently discussed as a candidate of dark matter \cite{Feng:2003nr,Dubovsky:2004ud,Pshirkov:2008nr,Aoki:2016zgp,Babichev:2016hir,Babichev:2016bxi,Aoki:2017cnz,Aoki:2017ffl,Marzola:2017lbt,Chu:2017msm,Albornoz:2017yup,Garny:2017kha}.

Massive bosonic fields can form self-gravitating objects which are called \emph{oscillatons} in a real massive scalar field \cite{Seidel:1991zh}, \emph{boson stars} in a complex massive scalar field~\cite{Kaup:1968zz,Jetzer:1991jr}, and \emph{Proca stars} in a complex massive vector field~\cite{Brito:2015pxa} (see also \cite{Schunck:2003kk,Herdeiro:2017fhv}). Hence, we can naturally expect that the massive graviton also yields self-gravitating solutions which we call \emph{massive graviton geons}. Indeed, the paper \cite{Aoki:2017cnz} showed the coherent oscillations of the massive graviton lead to the Jeans instability and then the massive gravitons could be self-gravitated. However, the analysis \cite{Aoki:2017cnz} is based on the linear perturbation theory around the homogeneous background. It has not been cleared whether or not the massive gravitons indeed yield a non-perturbative localized object.

We study non-trivial vacuum solutions to the bigravity theory. The bigravity theory contains a massless graviton and a massive graviton~\cite{Hassan:2011zd}. The energy-momentum tensor of the massive graviton was derived in \cite{Aoki:2016zgp,Aoki:2017cnz} in the similar way to general relativity~\cite{Isaacson:1968zza}. We find that the basic equations in the Newtonian limit can be reduced to the Schr\"{o}dinger-Poisson equations with the tensor ``wavefunction''. Due to the tensorial feature, the solutions in this system have different properties from the scalar case. We construct two types of the massive graviton geons; the monopole geon and the quadrupole geon. The monopole geon is the spherically symmetric configuration of the massive graviton which corresponds to the eigenstate of the zero \emph{total} angular momentum. On the other hand, the quadrupole geon is not spherically symmetric, instead, it is given by the quadrupole modes of the spherical harmonics. The quadrupole geon is the eigenstate of the zero \emph{orbital} angular momentum. Although the field configuration of the quadrupole geon is not spherical, it yields the spherically symmetric energy density and then the gravitational potential of the quadrupole geon is spherically symmetric. We also discuss the stability of the geons. The monopole geon is unstable against quadrupole mode perturbations. We then expect  a transition from the monopole geon  to the quadrupole one.
 The quadrupole geon would be a ground state of the massive graviton geons.

The paper is organized as follows. In Sec.~\ref{sec_Newtonian}, we introduce the bigravity theory and take the Newtonian limit with the self-gravity. We numerically construct the massive graviton geons in Sec.~\ref{sec_geons}. We find that not only the spherically symmetric configuration of the massive graviton but also a non-spherically symmetric configuration yields spherically symmetric energy density distributions. The perturbative stability of the monopole geon is studied in Sec.~\ref{sec_perturbation}. We give a summary and discuss some physical aspects of the massive graviton geons in Sec.~\ref{summary}. In Appendix \ref{app_harmonics}, we summarize multipole expansions of the tensor field and the definitions of the angular momenta. We also find an octupole configuration of the massive graviton geon in Appendix \ref{sec_oct}.

\section{Newtonian limit of bigravity}
\label{sec_Newtonian}
We assume the existence of a massive graviton as well as a massless graviton. The gravitational action is given by \cite{Hassan:2011zd}
\begin{align}
 S &=\frac{1}{2 \kappa _g^2} \int d^4x \sqrt{-g}R(g)+ \frac{1}{2 \kappa _f^2}
 \int d^4x \sqrt{-f} \mathcal{R}(f) 
 \nn
&-\frac{m_G^2}{ \kappa ^2} \int d^4x \sqrt{-g} \mathscr{U}(g,f) \,,
\label{action}
\end{align}
where $g_{\mu\nu}$ and $f_{\mu\nu}$ are two dynamical metrics, and
$R(g)$ and $\mathcal{R}(f)$ are their Ricci scalars.
The parameters $\kappa_g^2$ and $\kappa_f^2$ are 
the corresponding gravitational constants, 
while $\kappa$ is defined by $\kappa^2=\kappa_g^2+\kappa_f^2$.
Just for simplicity, to admit the Minkowski spacetime as a vacuum solution,
we restrict the potential $\mathscr{U}$ as the form \cite{deRham:2010ik,deRham:2010kj}
\begin{align}
\mathscr{U}&=\sum_{i=2}^4 c_i\mathscr{U}_i(\mathcal{K})\,,
\label{dRGT_potential}
\\
\mathscr{U}_2(\mathcal{K})&=-\frac{1}{4}\epsilon_{\mu\nu\rho\sigma} 
\epsilon^{\alpha\beta\rho\sigma}
{\mathcal{K}^{\mu}}_{\alpha}{\mathcal{K} ^{\nu}}_{\beta}\,, \nn
\mathscr{U}_3(\mathcal{K})&=-\frac{1}{3!}\epsilon_{\mu\nu\rho\sigma} 
\epsilon^{\alpha\beta\gamma\sigma}
{\mathcal{K} ^{\mu}}_{\alpha}{\mathcal{K} ^{\nu}}_{\beta}{\mathcal{K} ^{\rho}}_{\gamma}\,, 
\\
\mathscr{U}_4(\mathcal{K})&=-\frac{1}{4!}\epsilon_{\mu\nu\rho\sigma} 
\epsilon^{\alpha\beta\gamma\delta}
{\mathcal{K} ^{\mu}}_{\alpha}{\mathcal{K} ^{\nu}}_{\beta}{\mathcal{K} ^{\rho}}_{\gamma}
{\mathcal{K}^{\sigma}}_{\delta}\,,
\nonumber
\end{align}
with
\begin{align}
\mathcal{K}^{\mu}{}_{\nu}=\delta^{\mu}{}_{\nu}
-\left(\sqrt{g^{-1}f}\right)^{\mu}{}_{\nu}\,,
\end{align}
where $\left(\sqrt{g^{-1}f}\right)^{\mu}{}_{\nu}$ is defined by the relation
\begin{align}
\left(\sqrt{g^{-1}f}\right)^{\mu}{}_{\rho}
\left(\sqrt{g^{-1}f}\right)^{\rho}{}_{\nu}
=g^{\mu\rho}f_{\rho\nu}\,.
\end{align}
We can set $c_2=-1$ by using the normalization of the parameter $m_G$.
Then $g_{\mu\nu}=f_{\mu\nu}=\eta_{\mu\nu}$ is a vacuum solution of the bigravity,
and the parameter $m_G$ describes the mass of
the massive graviton propagating on the Minkowski background.

We assume that the curvature scales of the spacetimes are smaller than the graviton mass
\begin{align}
|\partial^2 g_{\mu\nu}| \ll m_G^2 \,, \quad
|\partial^2 f_{\mu\nu}| \ll m_G^2 \,. \label{small_curvature}
\end{align}
This means that we consider only weak gravity limit\footnote{The inequalities \eqref{small_curvature} also lead to that we do not need to take into account the Vainshtein mechanism~\cite{Vainshtein:1972sx}. For instance, the exterior region of a localized object with a mass $M$ yields $|\partial^2 g_{\mu\nu}| \sim GM/r^3$ and then \eqref{small_curvature} suggests the region outside the Vainshtein radius $r_V:=(GM/m_G^2)^{1/3}$.}. 

We perform the perturbations 
up to second orders around Minkowski spacetime and discuss 
a bound state of perturbed gravitational field.
The perturbed gravitational fields ($\delta g_{\mu\nu}$ and $\delta f_{\mu\nu}$) are rewritten into two modes, that is, 
\bea
&{\rm massless~mode:}&~~ \frac{1}{1+\alpha}\left(\delta g_{\mu\nu}+ \alpha \delta f_{\mu\nu}\right)
\,,
\\
&{\rm massive~mode:}&~~ \frac{\alpha}{1+\alpha} \left(\delta g_{\mu\nu}-\delta f_{\mu\nu}\right)
\,.
\ena
where $\alpha:=\kappa_g^2/\kappa_f^2$.
Those two modes are split into four components by those frequencies: 
the low-frequency massless mode $\delta\gB{}_{\mu\nu}$, the low-frequency massive mode $M_{\mu\nu}$, the high-frequency massless mode $h_{\mu\nu}$, and the high-frequency massive mode $\varphi_{\mu\nu}$, i.e., 
\bea
&{\rm massless~mode:}&~~\delta\gB{}_{\mu\nu}+h_{\mu\nu}/M_{\rm pl}
\\
&{\rm massive~mode:}&~~M_{\mu\nu}+\varphi^{\mu\nu}/M_G
\,.
\ena
where 
\begin{align}
M_{\rm pl}&:=\frac{\kappa}{\kappa_g\kappa_f}\,,\quad
M_G:=\frac{\kappa}{\kappa_g^2}\,.
\end{align}
Here the high frequency means that the frequency is same or larger than the graviton mass $m_G$, while the low frequency is that much smaller than $m_G$.
Intuitively, $\delta\gB{}_{\mu\nu}$ and $M_{\mu\nu}$ represent the Newtonian potential and the Yukawa potential, respectively. While, the high-frequency modes $h_{\mu\nu}$ and $\varphi_{\mu\nu}$ are the  propagating massless and massive spin-2 fields (see \cite{Aoki:2017cnz} for more details).

We shall focus on the scale beyond the Compton wavelength of the massive graviton ($m_G^{-1}$). In this scale, $M_{\mu\nu}$ can be ignored due to the Yukawa suppression. Since we are interested in a localized object composed of the massive graviton $\varphi_{\mu\nu}$, we assume the massless graviton $h_{\mu\nu}$ does not appear. As a result, the metrics $g_{\mu\nu}$ and $f_{\mu\nu}$ are approximated by 
\begin{align}
g_{\mu\nu}&\simeq \gB{}_{\mu\nu}+\frac{\varphi_{\mu\nu}}{M_G}
\,, \label{g_metric} \\
f_{\mu\nu}&\simeq \gB{}_{\mu\nu}-\frac{\varphi_{\mu\nu}}{\alpha M_G}
\,, \label{f_metric}
\end{align}
with $|\varphi_{\mu\nu}|/M_G \ll |\gB{}_{\mu\nu}|$ where $\gB{}_{\mu\nu}=\eta_{\mu\nu}+\delta\gB{}_{\mu\nu}$.

We then consider the perturbations up to second orders.
Just the similar to the GR case\cite{Isaacson:1968zza}, 
the high frequency modes $\varphi_{\mu\nu}$ 
give
the source for the low frequency gravitational potential, 
when we take the Isaacson average.
The low-frequency projection (the Isaacson average) is usually chosen as the spatial average (or the spacetime average) because 
of massless fields. However, here we consider the massive graviton, which momentum is much small compared with the rest mass energy $m_G$. 
Hence we perform the time average over the time interval $T=2\pi/m_G$ and we do not take any spatial average.

Under this setting, the Einstein equations in bigravity without matter fluids are reduced into
the Einstein-Klein-Gordon equations
\begin{align}
\GB{}^{\mu\nu}&=\frac{1}{M_{\rm pl}^2} \langle T_G^{\mu\nu} \rangle_{\rm low}\,, \label{Einstein_eq} \\
\left( \nablaB{}_{\alpha} \nablaB{}^{\alpha}-m_G^2 \right)\varphi_{\mu\nu}&=0+\mathcal{O}(\varphi_{\mu\nu}^2)\,, \label{KG_eq}
\end{align}
with the constraints
\begin{align}
\nablaB_{\mu}\varphi^{\mu\nu}=0+\mathcal{O}(\varphi_{\mu\nu}^2)\,, \quad 
\varphi^{\mu}{}_{\mu}=0+\mathcal{O}(\varphi_{\mu\nu}^2)\,, \label{constraints}
\end{align}
where we have ignored the higher order corrections of $\varphi_{\mu\nu}$ to the Einstein-Klein-Gordon equations. The Einstein tensor $\GB{}_{\mu\nu}$ is constructed by the low-frequency background $\gB{}_{\mu\nu}$ while $T_G^{\mu\nu}$ is the energy-momentum tensor of the high-frequency perturbations $\varphi_{\mu\nu}$. The energy-momentum tensor of massive graviton is defined by 
\begin{align}
T_{G}^{\mu\nu}&= -\left( \gB{}^{\mu\alpha} \gB{}^{\nu\beta}-\frac{1}{2}\gB{}^{\mu\nu} \gB{}^{\alpha\beta} \right) 
 \delta \Rff{}_{\alpha\beta}[\varphi] 
 \nn
&-\frac{m_G^2}{8} \left( 4 \varphi^{\mu\alpha} \varphi^{\nu}{}_{\alpha} - \gB{}^{\mu\nu} \varphi^{\alpha\beta} \varphi_{\alpha\beta} \right)+\mathcal{O}(\varphi_{\mu\nu}^3)\,,
\label{def_TG}
\end{align}
with
\begin{align}
&\delta \Rff{}_{\mu\nu}[\varphi]\nn
&= 
\Biggl[ \frac{1}{4} \nablaB{}_{\mu} \varphi^{\alpha \beta} \nablaB{}_{\nu} \varphi_{\alpha \beta}
+\nablaB{}^{\alpha} \varphi^{\beta}{}_{\nu} \nablaB{}_{[\alpha} \varphi_{\beta ] \mu}
\nn
&+\frac{1}{2}\varphi^{\alpha\beta} \left( \nablaB{}_{\nu} \nablaB{}_{\mu} \varphi_{\alpha\beta} +\nablaB{}_{\alpha} \nablaB{}_{\beta} \varphi_{\mu\nu} -2\nablaB{}_{\alpha} \nablaB{}_{(\mu} \varphi_{\nu) \beta} \right)
\nn
&
+\left( \frac{1}{2} \nablaB{}^{\beta} \varphi^{\alpha}{}_{\alpha} -\nablaB{}_{\alpha} \varphi^{\alpha\beta} \right)
\left( \nablaB{}_{(\mu}\varphi_{\nu) \beta} -\frac{1}{2}\nablaB{}_{\beta} \varphi_{\mu\nu} \right) \Biggl]\,.
\end{align}
The time average of this  energy-momentum tensor gives the source of the low-frequency background $\gB{}_{\mu\nu}$.
We have used the notations such that the suffixes on $\varphi_{\mu\nu}$ are raised and lowered by $\gB{}_{\mu\nu}$ and $\nablaB{}_{\mu}$ is the covariant derivative with respect to $\gB{}_{\mu\nu}$.

We then take the Newtonian limit: the non-relativistic limit of the massive graviton as well as the weak gravitational field approximation as for the background. The background metric is given by
\begin{align}
\gB{}_{\mu\nu}dx^{\mu}dx^{\nu}=-(1+2\Phi)dt^2+(1+2\Psi)\gamma_{ij}dx^i dx^j\,,
\end{align}
whereas the massive graviton is expressed by
\begin{align}
\varphi_{\mu\nu}=
\begin{pmatrix}
\psi_{00} & \psi_{0i} \\
* & \frac{\psi_{\rm tr}}{3} \gamma_{ij}+\psi_{ij} 
\end{pmatrix}
\frac{1}{\sqrt{2}} e^{-im_G t}+{\rm c.c.}\,,
\end{align}
where $\{\Phi,\Psi,\psi_{00},\psi_{0i},\psi_{\rm tr},\psi_{ij}\}$ are slowly varying functions of $(t,x^i)$, e.g. $\partial^2_t \psi_{ij} \ll m_G \partial_t \psi_{ij}$, and $\psi_{ij}$ is traceless, $\psi^{i}{}_i=0$. The three-dimensional Euclidean metric is denoted by $\gamma_{ij}$ and $\partial_i$ is the derivative with respect to $\gamma_{ij}$. We shall retain the three-dimensional covariance thus $\gamma_{ij}$ and $\partial_i$ are not necessary to be $\delta_{ij}$ and the partial derivatives, respectively. The indices $i,j$ are raised and lowered by $\gamma_{ij}$. The constraint equations \eqref{constraints} yield
\begin{align}
\psi_{\rm tr}=\psi_{00}\,,\quad \psi_{00}=\frac{i}{m_G} \partial^i \psi_{i0}\,,\quad
\psi_{0i}=\frac{i}{m_G}\partial^j \psi_{ij}\,,
\end{align}
which indicate the inequalities
\begin{align}
|\psi_{00}|,|\psi_{\rm tr}| \ll |\psi_{0i}| \ll |\psi_{ij}|\,.
\end{align}
Using the time average $\langle \cdots \rangle_{\rm low}$, we obtain
\begin{align}
\langle T^{\mu\nu}_G \rangle_{\rm low}\simeq {\rm diag}\left[ \frac{m_G^2}{4}\psi^*_{ij}\psi^{ij},0,0,0 \right]\,,
\end{align}
where $^*$ denotes the complex conjugate.
The equations are reduced to the Poisson-Schr\"{o}dinger equations
\begin{align}
\Delta \Phi&=\frac{m_G^2}{8M_{\rm pl}^2}\psi^*_{ij}\psi^{ij} \label{Poisson_eq}
\,, \\
i\frac{\partial}{\partial t} \psi_{ij}&=\left(
-\frac{\Delta}{2m_G}+m_G\Phi \right) \psi_{ij}\,, 
\label{Schrodinger_eq}
\end{align}
where $\Delta=\partial_i \partial^i$. The spatial component of the Einstein equation leads to $\Psi=-\Phi$. 
Note that these equations 
(\ref{Poisson_eq}) and (\ref{Schrodinger_eq}) are invariant under the rescaling
\begin{align}
\Phi \rightarrow \lambda^2\Phi\,,\quad \psi_{ij} \rightarrow \lambda^2 \psi_{ij}\,, \quad |x^i| \rightarrow \lambda^{-1} |x^i|\,, \quad  t\rightarrow \lambda^{-2}t\,, \label{rescale}
\end{align}
thus, the system is scale invariant. We also note that the ``wavefunction'' of the Schr\"{o}dinger equation is a tensor field $\psi_{ij}$ which yields differences from the scalar wavefunction case as we will see in the following sections.


\section{Self-gravitating massive gravitons}
\label{sec_geons}
In this section, we study the periodical solutions with the localized massive gravitons, i.e., the bound states of the system \eqref{Poisson_eq} and \eqref{Schrodinger_eq}. To find the bound states, we focus on some eigenstates of the angular momentum of $\psi_{ij}$. As detailed in Appendix \ref{app_harmonics}, the angular dependence of each eigenstate is given by the so-called pure-orbital spherical harmonics $(T^s_{j,j_z})_{ij}$ where $j$ and $j_z$ inside the parentheses are the total angular momentum quantum number and the total angular momentum projection quantum number. The total angular momentum $j$ and the orbital angular momentum $\ell$ are related by $j=\ell+s$ with $s=0,\pm 1, \pm 2$. We note that the label $s$ does not refer to polarizations of the massive graviton. The polarization eigenstates are explicitly shown in Appendix \ref{app_harmonics}.

Note that the Laplace operator $\Delta$ is connected with the orbital angular momentum $\ell$ whereas the symmetry of the field configuration is determined by $j$ and $j_z$. For instance, the spherically symmetric configuration is given by $j=j_z=0$ but this mode does not correspond to the zero orbital angular momentum mode as shown just below.

In what follows, we shall find two types of the massive graviton geons: the monopole geon and the quadrupole geon which correspond to the zero \emph{total} angular momentum mode $j=0$ and the zero \emph{orbital} angular momentum mode $\ell=0$, respectively. We discuss them in order. We will summarize their properties in Table \ref{table_geon}.

\subsection{Monopole geons}
\label{sec_monopole}
We first consider spherically symmetric bound states of Eqs.~\eqref{Poisson_eq} and \eqref{Schrodinger_eq} which we call the monopole geon. The spherical symmetry leads to
\begin{align}
\psi_{ij}dx^idx^j&=\sqrt{16\pi}\psi_0(r) e^{-iEt} (T^{-2}_{0,0})_{ij}dx^i dx^j
\nn
&=
\sqrt{\frac{2}{3}}\psi_0(r)e^{-iEt} \left(2dr^2
-r^2d\Omega^2\right)\,,
\end{align}
where $E$ is the ``energy eigenvalue'' of the bound state. Since we are interested in the bound state, $E$ is a real constant and $\psi_0$ is a real function of $r$. The Poisson-Schr\"{o}dinger equations are then given by
\begin{align}
\frac{1}{r^2}\frac{d}{dr}\left(r^2\frac{d\Phi}{dr}\right)&=4\pi G m_G^2 \psi_0^2 \,,
\end{align}
and
\begin{align}
\frac{1}{ r^2}\frac{d}{dr} \left(r^2\frac{d\psi_0}{dr}\right)=2\left( \frac{3 }{ r^2} +m_G^2 \Phi \right)\psi_0 &-2 m_G E\psi_0 \,, \label{Sch_ell=0}
\end{align}
where $G=1/8\pi M_{\rm pl}^2$.
The mass of the geon is defined by
\begin{align}
M:=4 \pi \int_0^{\infty} dr  r^2 m_G^2 \psi_0^2\,.
\end{align}
The mass $M$  is rescaled under the scaling \eqref{rescale} as
\begin{align}
M \rightarrow \lambda M \,.
\end{align}

The asymptotic forms of the solutions are analytically found as
\begin{align}
\Phi &\rightarrow -\frac{GM}{r} \,,  \\
\psi_0 &\rightarrow C r^{-1}W_{(-2\tilde{E})^{-1/2} ,\frac{5}{2}} \left( 2\sqrt{-2\tilde{E}}\tilde{r} \right) \nn
&\rightarrow C' \frac{e^{-\sqrt{-2\tilde{E}} \tilde{r}}}{r^{1-1/\sqrt{-2\tilde{E}}}}\,,
\end{align}
with
\begin{align}
\tilde{E}=\frac{E}{(GM)^2m_G^3}\,, \quad \tilde{r}=GMm_G^2 r \,, \label{rescaled_Er}
\end{align}
where $W_{\kappa,\mu}(z)$ is the Whittaker function, and $C$ and $C'$ are integration constants. We notice that the dimensionless variables $\tilde{E}$ and $\tilde{r}$ are invariant under the scaling \eqref{rescale}.

\begin{figure}[tbp]
\centering
\includegraphics[width=6cm,angle=0,clip]{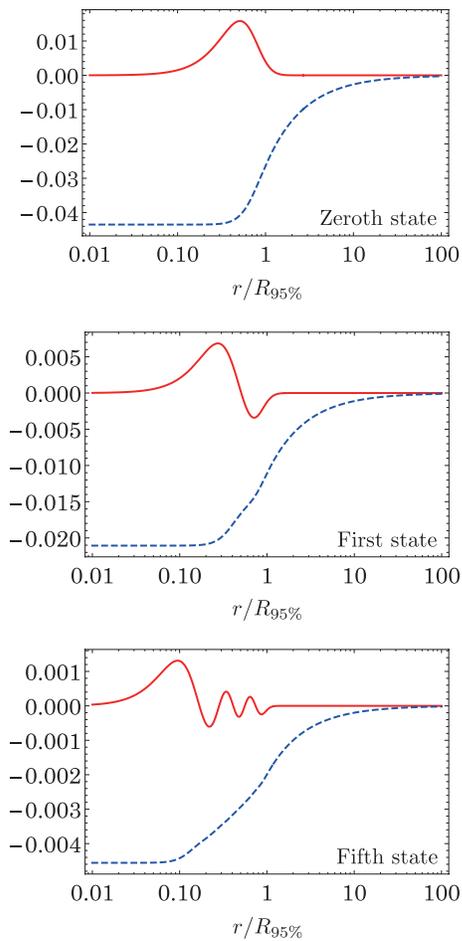}
\caption{The solutions to the Poisson-Schr\"{o}dinger equations of the spherically symmetric bound state $j=0$. The red curves show the configuration of the massive graviton $\tilde{\psi}_0$ and the blue dashed curves correspond to the gravitational potential $\tilde{\Phi}$. The solutions are specified by the number of nodes of $\tilde{\psi}_0$.
}
\label{fig_monopole}
\end{figure}

\begin{table}[tb]
\caption{Energy eigenvalues and $95\%$-mass radii of monopole geon.}
\label{table_ell=0}
\begin{tabular}{ccccc}
\hline
$\quad $   node     $\quad $      & $\tilde{E}$ & $\quad  \tilde{R}_{95\%} \quad $ \\
\hline\hline
$n=0$ & $-0.02708 $ & $37.72$ \\
$n=1$ & $-0.01140$ & $89.85$ \\
$n=2$ & $-0.006302$ & $161.9$ \\
$n=3$ & $-0.003995$ & $254.3$ \\
$n=4$ & $-0.002757$ & $367.4$ \\
$n=5$ & $-0.002016$ & $500.0$ \\
\hline
\end{tabular}
\end{table}

Some numerical solutions to the Poisson-Schr\"{o}dinger equations are shown in Fig.~\ref{fig_monopole} where we have used the dimensionless scale-invariant variables;
\begin{align}
\tilde{\Phi} = \frac{\Phi}{(GMm_G)^2} \,,  \quad
\tilde{\psi}_0 = \frac{\psi_0/M_{\rm pl}}{(GMm_G)^2 } \,.
\end{align}
The boundary condition is assumed to be $d\Phi/dr,$ $d\psi_0/dr \rightarrow 0$ as $r\rightarrow 0$, and $\Phi$ and $\psi_0$ vanish at infinity.
The solutions are parametrized by the number of nodes. We call the solution with $n$ nodes the $n$-th eigenstate of the monopole geon because the energy eigenvalue increases as $n$ increases as shown in Table \ref{table_ell=0}. In Table \ref{table_ell=0}, we summarize the scale-invariant energy eigenvalues $\tilde{E}$ and the $95\%$-mass radii $\tilde{R}_{95\%}$,
below which $95\%$ of the mass  energy is included. Since the $95\%$-mass radius increases in the higher-node state,
the lowest-node (the lowest-energy) state gives the most compact object.

The orbital angular momentum $\ell$ does not vanish in the $j=0$ mode since the zero total angular momentum in the tensor field is realized only when $\ell=+2$ and $s=-2$. As a result, the orbital angular momentum term $3/m_Gr^2$ exists in Eq. \eqref{Sch_ell=0}. The regularity at the center of the monopole geon leads to that $\psi_0$ vanishes at $r=0$. As a result, the monopole geon has a shell-like field configuration
as shown in Fig.~\ref{fig_monopole}. 
On the other hand, the Newtonian potential $\Phi$ is a scalar field, which has no orbital angular momentum for a spherical configuration, and then  $\Phi$
is finite at the center.


\subsection{Quadrupole geons}
\label{sec_quadrupole}
Next we 
 consider a non-spherically symmetric configuration of 
 the massive graviton, and find its bound state. 
 We assume a quadrupole configuration of $\psi_{ij}$, which
 is given by
\begin{align}
\psi_{ij}=\sqrt{16\pi}\sum_{j_z=-2}^{2}\psi_{2,j_z}(r) e^{-iEt} (T^{+2}_{2,j_z})_{ij} \,, \label{quadru}
\end{align}
where $(T^{+2}_{2,j_z})_{ij}$ is given by the spherical harmonics of the quadrupole mode. We call this bound state \eqref{quadru} the quadrupole geon. As shown in Appendix \ref{app_harmonics}, $(T^{+2}_{2,j_z})_{ij}$ has special properties which lead to
\begin{align}
\psi^*_{ij}\psi^{ij}&=4 \sum_{j_z=-2}^2 \psi_{2,j_z}^2 \,, \\
\Delta \psi_{ij}&= e^{-iEt}  \sum_{j_z=-2}^{2} \frac{\sqrt{16\pi}}{r^2}\frac{d}{dr} \left(r^2\frac{d\psi_{2,{j_z}} }{dr} \right) (T^{+2}_{2,j_z})_{ij} \,.
\end{align}
As a result, the energy density of $\psi_{ij}$ turns to be spherically symmetric. We then assume $\Phi=\Phi(r)$.
The basic equations are now given by
\begin{align}
\frac{1}{r^2}\frac{d}{dr}\left(r^2\frac{d\Phi }{dr}\right)
=4\pi G m_G^2 \sum_{j_z=-2}^2 \psi_{2,j_z}^2
\,,\\
\frac{1}{ r^2} \frac{d}{dr} \left(r^2\frac{d\psi_{2,j_z} }{dr} 
\right)=2m_G\left(m_G\Phi - E \right)\psi_{2,j_z} \,. \label{schro_ell2}
\end{align}
The orbital angular momentum $\ell$ vanishes in the quadrupole geon although that is not the case in the monopole geon. 
Note that the orbital angular momentum vanishes only when $j=2$ and $s=+2$.


\begin{table}[tb]
\caption{Energy eigenvalues and $95\%$-mass radii of quadrupole geon.}
\label{table_ell=2_one}
\begin{tabular}{ccccc}
\hline
$\quad$  node  $\quad$ & $\tilde{E}$ & $\quad  \tilde{R}_{95\%} \quad $ \\
\hline\hline
$n=0$ & $ -0.1628 \,$ & $7.830$ \\
$n=1$ & $-0.03080$ & $36.00$ \\
$n=2$ & $-0.01250$ & $85.04$ \\
$n=3$ & $-0.006747$ & $154.4$ \\
\hline
\end{tabular}
\end{table}

\begin{figure}[tbp]
\centering
\includegraphics[width=6cm,angle=0,clip]{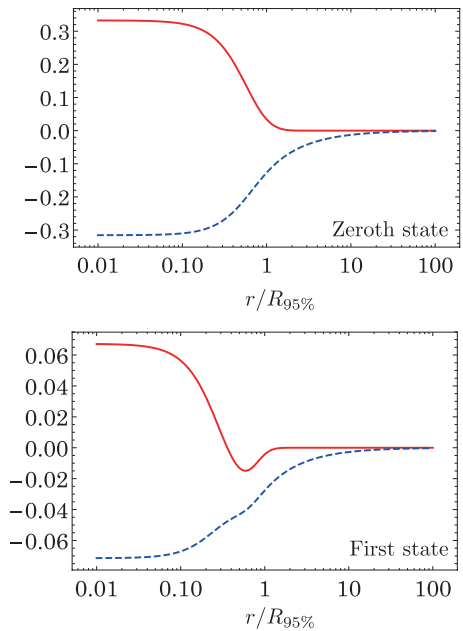}
\caption{The configurations of $\psi_2$ (red curves) and $\Phi$ (blue dashed curves) in the case of the quadrupole mode. 
}
\label{fig_qua_one}
\end{figure}

The variables $\psi_{2,j_z}$ are the solutions to the Sturm-Liouville equations with the same eigenvalue $E$ under the boundary conditions $d\psi_{2,j_z}/dr \rightarrow 0$ at the center and $\psi_{2,j_z}\rightarrow 0$ at infinity. Hence, the functional forms of $\psi_{2,j_z}$ are uniquely determined by some  function $\psi_{2E}(r)$ for each $E$ and then $\psi_{2,j_z}$ are expressed by
\begin{align}
\psi_{2,j_z}=a_{j_z} \psi_{2E} \,,
\end{align}
where $a_{j_z}$ are constants which are normalized as
\begin{align}
\sum_{j_z} |a_{j_z}|^2 =1 \,.
\label{constraint_aj}
\end{align}
Then, the equations are reduced to
\begin{align}
\frac{1}{r^2}\frac{d}{dr}\left(r^2\frac{d\Phi }{dr}\right)=4\pi G m_G^2 \psi_{2E}^2 \label{Poisson_ell2}
\,,\\
\frac{1}{r^2} \frac{d}{dr} \left(r^2\frac{d\psi_{2E} }{dr} \right)
=2m_G(m_G \Phi -E) \psi_{2E} \,, \label{Schro_ell2_simplify}
\end{align}
which are exactly the same equations of the Poisson-Schr\"{o}dinger system for the scalar field  with the $j=0$ mode.

The solutions to Eqs.~\eqref{Poisson_ell2} and \eqref{Schro_ell2_simplify} have been already investigated, e.g. in \cite{Ruffini:1969qy,bernstein1998eigenstates,Moroz:1998dh,harrison2002numerical,Guzman:2004wj,Hui:2016ltb}. The energy eigenvalues and the explicit forms of $\psi_{2E}$ are shown in Table.~\ref{table_ell=2_one} and Fig.~\ref{fig_qua_one}. Compared to those of the monopole geon, the lowest energy eigenvalue of the quadrupole geon ($\tilde E=-0.1628$) is much lower than that of the monopole geon ($\tilde E=-0.0278$). 
Since the lower energy state should be more stable than the higher state, the monopole geon may transit to the quadrupole geon. In next section, we will indeed find that the monopole geon is unstable against quadrupole mode perturbations.

Whereas the radial dependence of the quadrupole geon is uniquely determined by $\psi_{2E}$ for each $E$, the angular dependence is not unique because the expansion coefficients $a_{j_z}$  of  five tensor spherical harmonics $(T^{+2}_{2,j_z})_{ij}$ are arbitrary
although there is one constraint (\ref{constraint_aj}). 
 Each energy eigenstate of the quadrupole geon is degenerated. Needless to say, the scalar bosonic field with $j=0$ has no degeneracy. Although the equations \eqref{Poisson_ell2} and \eqref{Schro_ell2_simplify} are the same, there are different features between the gravitons and the scalar bosons.

Another difference from the bosonic case is that the field configuration is not spherically symmetric. The quadrupole geons have anisotropic pressures although it can be ignored to discuss the bound states. As a result, the quadrupole geons may emit gravitational waves. We will return some consequences of the gravitational wave emission in Sec.~\ref{summary}.

Before ending this section, we comment on the polarizations of the geons. The massive graviton has spin-0, spin-1, and spin-2 polarization modes. The monopole geon is an eigenstate of the spin-0 polarization while the quadrupole geon is not an eigenstate of the polarizations. The relation between the polarization eigenstates and the angular momentum eigenstates is given by Eq.~\eqref{ortho} in Appendix \ref{app_harmonics} which explicitly shows that $(T^{+2}_{2,j_z})_{ij}$ is not the polarizations eigenstate.


\section{Linear fluctuations around monopole geon}
\label{sec_perturbation}
In the previous section, we find two types of geon solutions; monopole and quadrupole.  Since the binding energy ($-\tilde E$) of 
the quadrupole geon is much larger than that of the monopole one, 
the monopole geon is expected to be unstable.
In order to confirm  it, 
we study the perturbations around the monopole geon.
As long as the perturbations do not spoil the Newtonian approximation, we can use the equations \eqref{Poisson_eq} and \eqref{Schrodinger_eq}. Hence, we consider the configurations
\begin{align}
\Phi=\Phi_0+\delta \Phi \,, \quad
\psi_{ij}=\psi_{0,ij}+\delta \psi_{ij}\,,
\end{align}
where $\Phi_0$ and $\psi_{0,ij}$ are the solutions of the $n$-th state of the monopole geon. The perturbations $\delta \psi_{ij}$ can be decomposed into the odd parity perturbations and the even parity perturbations,
\begin{align}
\delta \psi_{ij}&=\delta \psi_{ij}^{\rm (even)}+\delta \psi_{ij}^{\rm (odd)} \,,
\end{align}
and the different parity modes are not coupled with each other.

There is no odd parity mode of the gravitational potential. The odd parity perturbations $\delta \psi^{\rm (odd)}_{ij}$ are obtained by only solving the Schr\"{o}dinger equation with the potential $\Phi=\Phi_0$ and then the problem can be reduced to the eigenvalue problem of a self-adjoint operator. Hence, the eigenvalues, i.e., the frequencies of $\delta \psi^{\rm (odd)}_{ij}$, are real and then there is no growing mode (and no decaying mode). The monopole geons are linearly stable against the odd parity perturbations.

We then study the even parity perturbations. We introduce a different basis of the multipole expansion form $(T^s_{j,j_z})_{ij}$ which is called the pure-spin spherical harmonics denoted by $(Y^A_{j,j_z})_{ij}$ with $A=S_0,E_1,E_2,B_1,B_2$. 
\footnote{We have used two types of the orthonormal sets of the tensor harmonics $(T^{s}_{j,j_z})_{ij}$ and $(Y^{A}_{j,j_z})_{ij}$ which are related by the orthogonal transformation \eqref{ortho}. Although $(T^{s}_{j,j_z})_{ij}$ are useful to discuss the bound state solutions, the perturbation equations are given by a simpler form by using $(Y^{A}_{j,j_z})_{ij}$.}
The label $A$ characterizes the polarization states of the massive spin-2 field and the parity (see Appendix \ref{app_harmonics}). Then, the even parity perturbations can be expressed by
\begin{align}
\delta \Phi&=\sqrt{4\pi} \sum_{j \geq0} \sum_{|j_z|\leq j} \delta \phi_{j,j_z}(t,r)  Y_{j,j_z}
\,, \label{phi_pert} \\
\delta \psi^{\rm (even)}_{ij}&=\sqrt{16\pi} e^{-iEt} \sum_{j \geq0} \sum_{|j_z|\leq j} \delta \xi_{j,j_z}(t,r) (Y^{S_0}_{j,j_z})_{ij} 
\nn
&+\sqrt{16\pi} e^{-iEt} \sum_{j \geq 1} \sum_{|j_z|\leq j} \delta  \chi_{j,j_z}(t,r) (Y^{E_1}_{j.j_z})_{ij} 
\nn
&+\sqrt{16\pi} e^{-iEt} \sum_{j \geq 2} \sum_{|j_z|\leq j} \delta \sigma_{j,j_z}(t,r) (Y^{E_2}_{j,j_z})_{ij} \,,
\label{psi_pert}
\end{align}
where $Y_{j,j_z}$ is the scalar spherical harmonics. Since the different modes of $j$ and $j_z$ are not coupled due to the background spherical symmetry, we just omit the suffixes $j$ and $j_z$ hereafter.

The perturbed Poisson equation yields
\begin{align}
\frac{1}{r}\frac{\partial^2}{\partial r^2}(r\delta \phi)-\frac{j(j+1)}{r^2}\delta \phi=4\pi G \psi_0 (\delta \xi+\delta \xi^* )\,,
\end{align}
while the Schr\"{o}dinger equation with $j \geq 2$ gives
\begin{align}
-\frac{1}{2m_Gr}\frac{\partial^2}{\partial r^2}(r\delta \xi)+\left[\frac{j(j+1)+6}{2m_G r^2} +m_G U \right]&\delta \xi 
\nn
+m_G \psi_0 \delta \phi+\frac{\sqrt{3j(j+1)}}{m_Gr^2}\delta\chi
&=i \frac{\partial}{\partial t}\delta \xi
, \label{eq_psi} \\
-\frac{1}{2m_G r}\frac{\partial^2}{\partial r^2}(r\delta \chi)+\left[\frac{j(j+1)+4}{2m_G r^2}+m_G U\right]&\delta \chi \nn
+\frac{\sqrt{3j(j+1)}}{m_G r^2}\delta \xi+\frac{\sqrt{(j-1)(j+2)}}{m_G r^2}\delta \sigma&=i\frac{\partial}{\partial t} \delta \chi 
, \label{eq_chi} \\
-\frac{1}{2m_Gr}\frac{\partial^2}{\partial r^2}(r\delta \sigma) + \left[ \frac{(j-1)(j+2)}{2m_Gr^2}+m_G U\right]& \delta \sigma \nn
+\frac{\sqrt{(j-1)(j+2)}}{m_Gr^2}\delta \chi&=i \frac{\partial}{\partial t}\delta \sigma , \label{eq_sigma}
\end{align}
where $U=\Phi_0-E/m$. We note $j=0$ and $j=1$ of the perturbations are exceptional modes. The variables $\delta \chi$ and $\delta \sigma$ are undefined for the $j=0$ mode in which \eqref{eq_chi} and \eqref{eq_sigma} are not obtained. For the $j=1$ mode, the variable $\delta \sigma$ is undefined and the equation \eqref{eq_sigma} do not exist. However, Eq.~\eqref{eq_psi} (and Eq.~\eqref{eq_chi}) is correct even for the $j=0$ mode (and the $j=1$ mode) since the coefficients in front of the undefined variable vanish.

We find solutions of the form
\begin{align}
\delta \phi&=\frac{W}{r}e^{-i {\omega} t}+\frac{W^*}{r} e^{i{\omega}^* t} \,, \label{W_def} \\
\delta \xi &=\frac{1}{r}( \xi_A+\xi_B)e^{-i {\omega} t}+\frac{1}{r}(\xi_A^*- \xi_B^*)e^{i{\omega}^* t} \,, \\
\delta\chi &=\frac{1}{r}(\chi_A+\chi_B)e^{-i {\omega} t}+\frac{1}{r}(\chi_A^*-\chi_B^*)e^{i {\omega}^* t} \,, \\
\delta \sigma &=\frac{1}{r}(\sigma_A+\sigma_B)e^{-i {\omega} t}+\frac{1}{r}(\sigma_A^*-\sigma_B^*)e^{i {\omega}^* t} \,, \label{sigma_def}
\end{align}
where $\{W, \xi_{A,B} ,\chi_{A,B} ,\sigma_{A,B} \}$ are complex functions of $r$ and ${\omega}$ is a complex constant. The regularity conditions at the center lead to the boundary condition $\{W, \xi_{A,B} ,\chi_{A,B} ,\sigma_{A,B} \}\rightarrow 0$. At infinity we assume the boundary condition $\delta \xi,\delta \psi, \delta \chi, \delta \sigma\rightarrow 0$. Except for the $j=0$ mode, all variables $\{\delta \phi,\delta \xi,\delta \chi,\delta \sigma\} $ decay faster than $r^{-1}$. Hence, we can assume the boundary condition $\{W, \xi_{A,B} ,\chi_{A,B} ,\sigma_{A,B} \}\rightarrow 0$ at infinity. However, $\delta \phi$ of the $j=0$ mode decays as $r^{-1}$, i.e., $W\rightarrow W_0$ (constant). We cannot assume $W\rightarrow 0$ at infinity in the case of $j=0$; instead, we should assume $dW/dr\rightarrow 0$. 

The Poisson equation yields that $W$ is formally given by
\begin{align}
W=-8\pi G \mathcal{O}_{\phi}^{-1} \psi_0 \xi_A \,,
\end{align}
where 
$\mathcal{O}_{\phi}^{-1}$ is the inverse operator of $\mathcal{O
}_{\phi}$, which is defined by
\begin{align}
\mathcal{O}_{\phi}&=-\frac{d^2}{dr^2}+\frac{j(j+1)}{r^2} \,.
\end{align}
 The perturbed equations for the $j=0$ mode is given by
\begin{align}
\begin{pmatrix}
0 & \mathcal{O}_{\psi} \\
\mathcal{O}_{\psi}-8\pi G m_G \psi_0 \mathcal{O}_{\phi}^{-1} \psi_0 & 0
\end{pmatrix}
\begin{pmatrix}
 \xi_A \\
 \xi_B
\end{pmatrix}
={\omega}
\begin{pmatrix}
\xi_A \\
 \xi_B
\end{pmatrix}
,
\end{align}
the equations for $j=1$ are
\begin{widetext}
\begin{align}
\begin{pmatrix}
0 & \mathcal{O}_{\psi} & 0 & \frac{\sqrt{6}}{m_G r^2}\\
\mathcal{O}_{\psi}-8\pi G m_G \psi_0 \mathcal{O}_{\phi}^{-1}\psi_0 & 0 & \frac{\sqrt{6}}{m_G r^2} & 0 \\
0 & \frac{\sqrt{6}}{m_G r^2} & 0 & \mathcal{O}_{\chi}  \\
\frac{\sqrt{6}}{m_G r^2} & 0 & \mathcal{O}_{\chi} & 0
\end{pmatrix}
\begin{pmatrix}
 \xi_A \\
\xi_B \\
\chi_A \\
\chi_B 
\end{pmatrix}
={\omega}
\begin{pmatrix}
 \xi_A \\
 \xi_B \\
\chi_A \\
\chi_B 
\end{pmatrix}
,
\end{align}
and the equations for the general modes $j\geq 2$ are written as
\begin{align}
\begin{pmatrix}
0 & \mathcal{O}_{\psi} & 0 & \frac{\sqrt{3j(j+1)}}{m_G r^2} & 0 & 0 \\
\mathcal{O}_{\psi}-8\pi G m_G \psi_0 \mathcal{O}_{\phi}^{-1}\psi_0 & 0 & \frac{\sqrt{3j(j+1)}}{m_G r^2} & 0 & 0 & 0 \\
0 & \frac{\sqrt{3j(j+1)}}{m_G r^2} & 0 & \mathcal{O}_{\chi} & 0 & \frac{\sqrt{(j-1)(j+2)}}{m_G r^2} \\
\frac{\sqrt{3j(j+1)}}{m_G r^2} & 0 & \mathcal{O}_{\chi} & 0 & \frac{\sqrt{(j-1)(j+2)}}{m_G r^2} & 0 \\
0 & 0 & 0 & \frac{\sqrt{(j-1)(j+2)}}{m_G r^2} & 0 & \mathcal{O}_{\sigma} \\
0 & 0 & \frac{\sqrt{(j-1)(j+2)}}{m_G r^2} & 0 & \mathcal{O}_{\sigma} & 0
\end{pmatrix}
\begin{pmatrix}
\xi_A \\
\xi_B \\
\chi_A \\
\chi_B \\
\sigma_A \\
\sigma_B 
\end{pmatrix}
={\omega}
\begin{pmatrix}
 \xi_A \\
\xi_B \\
\chi_A \\
\chi_B \\
\sigma_A \\
\sigma_B 
\end{pmatrix}
,
\end{align}
\end{widetext}
where
\begin{align}
\mathcal{O}_{\psi}&=-\frac{1}{2m_G }\frac{d^2}{dr^2}+\frac{j(j+1)+6}{2m_G r^2}+m_G U \,, \\
\mathcal{O}_{\chi}&=-\frac{1}{2m_G }\frac{d^2}{dr^2}+\frac{j(j+1)+4}{2m_G r^2}+m_G U \,, \\
\mathcal{O}_{\sigma}&=-\frac{1}{2m_G }\frac{d^2}{dr^2}+\frac{(j-1)(j+2)}{2m_G r^2}+m_G U \,.
\end{align}

Once we find one solution with  the eigenvalue ${\omega}$,
we can easily obtain the solutions with the eigenvalues $-{\omega},\omega^*$ and $-{\omega}^*$. Supposing $\{W,\xi_A,\ \xi_B,\cdots \}$ with ${\omega}$ is a solution, the solution with $-{\omega}$ is given by $\{W, \xi_A,- \xi_B,\cdots \}$. The solutions with eigenvalues ${\omega}^*$ and $-\omega^*$ are also easily constructed from the definitions of the variables \eqref{W_def}-\eqref{sigma_def}.

We use the spectral method and then the problem is reduced to the eigenvalue problem of the matrix. Since the infinity cannot be treated in numerical calculations, we choose a range $(0,10R_{95\%})$ for $r$ for which we have confirmed the numerical results are sufficiently converged. 
When all eigenvalues are real, the solution is stable, while if there exists an imaginary part of ${\omega}$, there are always exponentially growing modes because $\{ {\omega},-{\omega},{\omega}^*,-{\omega}^*\}$ are the eigenvalues of the perturbations.

In Tables \ref{table_pert1} and \ref{table_pert2}, we show some eigenvalues of ${\rm Re}({\omega})\geq 0, {\rm Im} (\omega)\geq 0$ where the rescaled ${\omega}$ is defined by
\begin{align}
\tilde{{\omega}}=\frac{{\omega}}{(GM)^2m_G^3} \,.
\end{align}
We note that the $j=0$ mode perturbations have a trivial solution such that ${\omega}=0$ and
\begin{align}
\xi_B \propto \psi_0/r\,, \quad  \xi_A=0\,, \quad \delta \phi=0\,,
\end{align}
which corresponds to a phase shift of the background solution, i.e., $\psi_{0,ij}+\delta \psi_{ij}=e^{i\epsilon} \psi_{0,ij}$ with a small constant $\epsilon$. We also find almost zero eigenvalues in the $j=1$ and $j=2$ modes perturbations which are not trivial solutions.

\begin{table}[tb]
\caption{The seven lowest eigenvalues $\tilde{{\omega}}_i$ of the $j=0$ mode perturbations with ${\rm Re}(\tilde{\omega}_i),{\rm Im}(\tilde{\omega}_i)\geq 0$ and $|\tilde{\omega}_0| < |\tilde{\omega}_1| < |\tilde{\omega}_2| <\cdots$.}
\label{table_pert1}
\begin{tabular}{ccccc}
\hline
& zeroth state  & first state & second state \\
\hline\hline
$\tilde{{\omega}}_0$ & 0       & 0 & 0\\
$\tilde{{\omega}}_1$ & 0.005127 & 0.001389 & 0.0005677\\
$\tilde{{\omega}}_2$ & 0.01326 & 0.003794+0.0007111$i$ & 0.001571+0.0003633$i$\\
$\tilde{{\omega}}_3$ & 0.01699 & 0.005369 & 0.002495 \\ 
$\tilde{{\omega}}_4$ & 0.01936 & 0.006537 & 0.002879+0.0001624$i$ \\
$\tilde{{\omega}}_5$ & 0.02095 & 0.007374 & 0.003445 \\
$\tilde{{\omega}}_6$ & 0.02197 & 0.008009 & 0.003839 \\
\hline
\end{tabular}
\end{table}

\begin{table}[tb]
\caption{The seven lowest eigenvalues $\tilde{{\omega}}_i$ of perturbations around the zeroth state of the monopole geon.}
\label{table_pert2}
\begin{tabular}{ccccc}
\hline
& $\qquad  j=1 \qquad $ & $\qquad j=2 \qquad $ & $\qquad  j=3 \qquad  $\\
\hline\hline
$\tilde{{\omega}}_0$ & 0.00000 & 0.00000 & 0.004440 \\
$\tilde{{\omega}}_1$ & 0.004674 &  0.0005155$i$ \, & 0.004918\,\\
$\tilde{{\omega}}_2$ & 0.00622 & 0.008190 & 0.005600 \\
$\tilde{{\omega}}_3$ & 0.01078 & 0.008469 & 0.01133 \\
$\tilde{{\omega}}_4$ & 0.01132 & 0.008660 & 0.01189 \\
$\tilde{{\omega}}_5$ & 0.01551 & 0.01070 & 0.01346 \\
$\tilde{{\omega}}_6$ & 0.01581 & 0.01358 & 0.01559 \\
\hline
\end{tabular}
\end{table}

\begin{figure}[tbp]
\centering
\includegraphics[width=6cm,angle=0,clip]{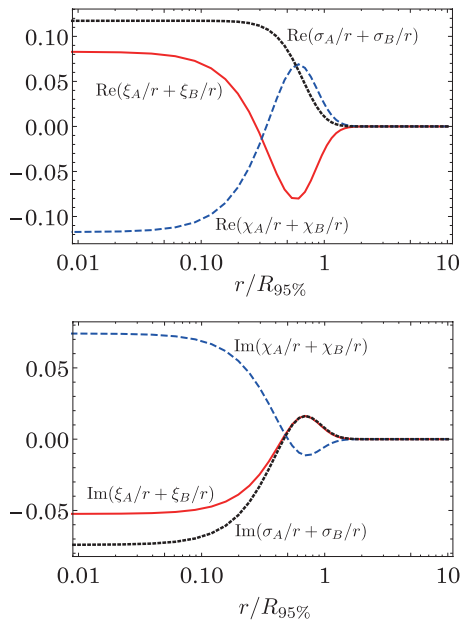}
\caption{The solution to the $j=2$ mode perturbation equations with $\tilde{{\omega}}=0.00052i$ (the unstable perturbation around the monopole geon). The amplitudes of the solutions are scale free because the equations are linear.
}
\label{fig_perturbation}
\end{figure}

We first discuss the $j=0$ mode perturbations. 
We find the $n$-th state has the $n$ different unstable modes as shown in Table \ref{table_pert1}. The unstable modes could be interpreted as the transition modes to the lower energy eigenstate from the higher one. Only the zeroth state of the monopole geon is linearly stable against the $j=0$ mode perturbation.

Henceforth, we shall only show the results of the perturbations around the zeroth state.
Our numerical calculations show that the zeroth state is linearly stable against the $j\neq 2$ mode perturbations. However, we find there exists an unstable mode in the $j=2$ mode perturbations as shown in Table \ref{table_pert2}. The explicit forms of the unstable mode are depicted in Fig.~\ref{fig_perturbation}.

As a result, the zeroth state of the monopole geon is unstable against the $j=2$ mode perturbation which is consistent with the energy argument. However, it is not so manifest whether the unstable mode indeed represents the transition to the zeroth state of the quadrupole geon from the zeroth state of the monopole geon. The quadrupole geon is purely given by the $j=2$ mode and the gravitational potential is exactly spherically symmetric. We expect that the $j\neq 2$ modes of $\psi_{ij}$ and the non-spherical pert of $\Phi$ decay in time and then the configurations of $\Phi$ and $\psi_{ij}$ will converge towards 
 the zeroth state of the quadrupole geon. However, we need to fully solve \eqref{Poisson_eq} and \eqref{Schrodinger_eq} 
 for the monopole geon with perturbations 
 in order to discuss this process, which we leave for a future work.

Furthermore, the stability of the quadrupole geon is an open question. Since the field configuration is not spherically symmetric in this case, different modes of the multipole expansion should be coupled and the stability analysis becomes very difficult. We hope
 that the quadrupole geon is linearly stable since it is the only mode with the zero orbital angular momentum and thus the lowest energy eigenstate of \eqref{Poisson_eq} and \eqref{Schrodinger_eq}.


\section{Summary and Discussions}
\label{summary}
We find vacuum solutions such that the massive gravitons are localized by their gravitational energy in the context of the bigravity theory, which we have called the massive graviton geon by following Wheeler's idea~\cite{Wheeler:1955zz}. In the Newtonian limit, the vacuum equations of the bigravity theory can be reduced to the Poisson-Schr\"{o}dinger equations with the tensor ``wavefunction''. We numerically construct two types of the massive graviton geons: the spherically symmetric configuration of the massive graviton (the monopole geon) and the configuration is represented by the quadrupole modes of the spherical harmonics (the quadrupole geon).
We summarize their properties in Table \ref{table_geon}. We find the energy density distribution is spherically symmetric not only for the monopole geon but also for the quadrupole geon. The lower energy eigenstate should be more stable than the higher eigenstate. Indeed, the perturbation analysis around the monopole geon reveals that there exists an unstable mode in the quadrupole mode perturbations which is consistent with the energy argument. 

\begin{table}[h]
\caption{Properties of the massive graviton geons.}
\label{table_geon}
\begin{tabular}{l|l||c|c}
\hline
\multicolumn{2}{l||}{~}& monopole geon & quadrupole geon\\
\hline\hline
\multicolumn{2}{l||}{angular momentum} & $j=0,\ell=2$ & $j=2,\ell=0$ 
\\
\hline
\raisebox{-2ex}{configuration} & angular & spherical& quadrupole
\\
& radial  & shell (Fig.~\ref{fig_monopole}) & ball (Fig.~\ref{fig_qua_one})
\\
\hline
\multicolumn{2}{l||}{lowest energy }& $\tilde{E}=-0.027$ & $\tilde{E}=-0.16$ \\
\hline
\multicolumn{2}{l||}{ degeneracy }& 1 & 5 \\
\hline
\multicolumn{2}{l||}{polarization} & scalar & not polarized
\\
\hline
\end{tabular}
\end{table}

Since the system is scale invariant, we can construct a massive graviton geon with any size of $R_{95\%}$ as long as $R_{95\%} \gg m_G^{-1}$. The most stable and compact massive graviton geon would be given by the zeroth state of the quadrupole geon whose mass and typical amplitude are
\begin{align}
GM &\simeq \frac{7.8}{m_G^2 R_{95\%}} \,, \label{geon_mass} \\
|\varphi_{\mu\nu}|/M_G &\simeq 0.3 \times \alpha^{1/2} (GMm_G)^2 \,. \label{geon_ampli}
\end{align}

If the size of the geon becomes smaller such as $R_{95\%} \sim m_G^{-1}$, we cannot ignore the relativistic effects and the present approximations are no longer valid. Nevertheless, 
we may evaluate the maximum mass of the massive graviton geon from the expression \eqref{geon_mass} as follows.
If $R_{95\%} \lsim m_G^{-1}$, $M\gsim (Gm_G)^{-1}$, and then 
the Schwarzschild radius $2GM$ becomes larger than the size of the massive graviton geon $R_{95\%} $, which may become a black hole in a relativistic situation. Then we expect that 
the massive graviton geon exists only when $M\lesssim M_{\rm max}$,
where 
\begin{align}
M_{\rm max} \sim (Gm_G)^{-1} \sim 1 M_{\odot} \left(\frac{10^{-10}{\rm eV}}{m_G} \right)
\,.
\end{align}
This mass bound of the geon is consistent with the stability analysis of the Schwarzschild black hole in the bigravity. The papers \cite{Babichev:2013una,Brito:2013wya} showed that the Schwarzschild black hole solution is stable when $2GM\gtrsim m_G^{-1}$ while it turns to be unstable when $2GM\lesssim m_G^{-1}$,  
in which case we may expect a hairy black hole surrounded by  
time-dependent massive gravitons  is formed.

The quadrupole geons may emit gravitational waves. We note, however, that the low-frequency part of $T^{\mu\nu}_G$ has no oscillation due to the phase cancellation up to the second order. Hence, 
we  study the backreaction  of the massive graviton $\varphi_{\mu\nu}$ on the high-frequency massless gravitons $h_{\mu\nu}$. 
We find  $ T_G^{00}$ is static and spherically symmetric, while 
the spatial components are given by
\begin{align}
 T_G^{ij} \simeq m_G^2 \left( - \psi^{ik} \psi^j{}_k +\frac{3}{4}\gamma^{ij} \psi^{kl}\psi_{kl} \right) e^{-2i m_G t} +{\rm c.c.}\,, \label{anisotropic_stress}
\end{align}
which clearly has the oscillating anisotropic stress and then the quadrupole geons emit the gravitational waves $h_{\mu\nu}$. The gravitational waves observed sufficiently far from the geon are given by
\begin{align}
h_{ij}^{\rm TT}=\frac{4G}{r}e^{i{\small \vect{k}}\cdot {\small \vect{x}}} P_{ij,kl} \int d^3 {\small \vect{x}}'  T_G^{kl}(t,{\small \vect{x}}') e^{-i{\small \vect{k}}\cdot {\small \vect{x}}'}+
{\rm c.c.}\,, \label{GWs}
\end{align}
where ${\small \vect{k}}$ is the wavevector of the gravitational waves with ${\small \vect{k}}^2=(2m_G)^2$ and $P_{ij,kl}$ is the transverse-traceless projection operator. Note that the pure-orbital spherical harmonics with $j=2$ and $s=+2$ are just constant matrices in the Cartesian coordinates (see Appendix \ref{app_harmonics}). Hence, the integration in \eqref{GWs} is 
\begin{align} 
\int d^3{\small \vect{x}}'  T_G^{kl} e^{-i{\small \vect{k}}\cdot {\small \vect{x}}' }
&\propto \int d^3 {\small \vect{x}}' \psi_{2E}^2(r') e^{-i{\small \vect{k}}\cdot {\small \vect{x}}'} \nn
&=\frac{2\pi}{m_G} \int dr' r' \psi_{2E}^2(r') \sin[2m_G r'] \,.
\end{align}
As a result, the gravitational wave emission is suppressed (probably exponentially) for the non-relativistic geons, i.e., $\partial_r \psi_{2E} \ll m_G \psi_{2E}$. The non-relativistic quadrupole geons are long-lived objects even if they have the oscillating part of the anisotropic stress.


The gravitational potential of the quadrupole geon has the same structure as the case of a massive scalar field of the $j=0$ mode because Eqs. \eqref{Poisson_ell2} and \eqref{Schro_ell2_simplify} are the same ones as those in the scalar case (see e.g., \cite{Ruffini:1969qy,bernstein1998eigenstates,Moroz:1998dh,harrison2002numerical,Guzman:2004wj,Hui:2016ltb}). Hence, the quadrupole geon can represent a central part of a dark matter halo if the mass of the graviton is about $10^{-22}$ eV just as the case with the ultralight scalar dark matter (see \cite{Hui:2016ltb} for example and references therein). 

Because the matter fields are coupled with $g_{\mu\nu}$ given by \eqref{g_metric}, the massive graviton $\varphi_{\mu\nu}$ can be observed as localized ``gravitational waves'' 
with the frequency and amplitude of 
\begin{align}
f&\simeq m_G/2\pi \sim 10^{-8} \left(\frac{m_G}{10^{-22}{\rm eV}} \right) {\rm Hz} \,, \\
\frac{|\varphi_{\mu\nu}|}{M_G} &\sim 10^{-7} \alpha^{1/2}  \left(\frac{M}{10^9 M_{\odot}} \right)^2
\,.
\end{align}
In the case $\alpha=(M_{\rm pl}/ M_G)^2 \sim 1$, i.e., $M_{\rm pl} \sim M_G$, this oscillation has a large amplitude thus such a possibility is already excluded by the pulsar timing array (PTA) observations \cite{Hobbs:2017oam}.
One possibility to be a consistent scenario is introducing the hierarchy of the mass scales $M_G \gtrsim 10^8 M_{\rm pl}$ ($\alpha^{1/2} \lesssim 10^{-8}$).
Another possibility is introducing the $Z_2$ symmetry of the massive graviton~\cite{Aoki:2017cnz}; the theory is symmetric under $\varphi_{\mu\nu} \rightarrow -\varphi_{\mu\nu}$. The symmetry leads to $M_G=M_{\rm pl}$ and the matter fields are coupled with the metric \cite{deRham:2014naa}
\begin{align}
g_{\rm eff}{}_{\mu\nu}&=\frac{1}{4}\left[ g_{\mu\nu}+2g_{\mu\alpha} \left( \sqrt{g^{-1}f}\right)^{\alpha}{}_{\nu}+f_{\mu\nu} \right] \nn
&= \gB{}_{\mu\nu}-\frac{1}{4M_{\rm pl}^2} \varphi_{\mu\alpha} \varphi^{\alpha}{}_{\nu}+\cdots \,. \label{eff_metric}
\end{align}
In this case, the amplitude of the oscillation is given by $|\varphi_{\mu\alpha} \varphi^{\alpha}{}_{\nu}/4M_{\rm pl}^2| \sim 10^{-15}$, which can be detectable by the future observations of PTA. We note that the ultralight scalar dark matter predicts oscillations of the gravitational potential because the coherent oscillation of dark matter leads to the oscillation of the pressure~\cite{Khmelnitsky:2013lxt}. In the tensor case, 
we also expect the oscillation of the gravitational potential (see \eqref{anisotropic_stress}). 

To discuss the dark matter scenario, we have to take into account other observational constraints, such as the Yukawa force constraints and the gravitational wave constraints, on the graviton mass. Supposing the linear theory is a good approximation in the observational scales, the mass range $10^{-23}{\rm eV} \lesssim m_G \lesssim 10^{-4} {\rm eV}$ is excluded when $M_{\rm pl} \sim M_G$~(see \cite{Will:2014kxa,Murata:2014nra,deRham:2016nuf,Baker:2017hug} for examples). However, if $M_G \gg M_{\rm pl}$, the interaction of the massive graviton becomes weak and then the constraints are relaxed (For instance, the lunar laser ranging experiment can give a graviton mass constraint only if $(M_{\rm pl}/ M_G)^2 \gtrsim 10^{-11}$~\cite{Merkowitz:2010kka}). Furthermore, the observational constraints cannot be applied to the $Z_2$ symmetric model since the $Z_2$ symmetry prohibits the Yukawa interaction $M_G^{-1} \varphi_{\mu\nu}T^{\mu\nu}$. As a result, the graviton mass constraints cannot be applied to the graviton dark matter models with the hierarchy $M_G \gtrsim 10^8 M_{\rm pl}$ or with the $Z_2$ symmetry.

The gravitational geons are recently discussed in the context of the nonlinear instability of the AdS spacetime~\cite{Bizon:2011gg,Dias:2012tq,Horowitz:2014hja,Martinon:2017uyo}. 
The energy eigenvalue of the massive graviton geon is negative which suggests that the massive graviton geons can be produced from fluctuations around the Minkowski vacuum. Does it imply the 
instability of Minkowski spacetime
if a graviton is massive? 
We expect that is not the case.
Because the mass energy of the produced geon is not negative although 
the energy eigenvalue is negative. 
When all fluctuation energy is converted into the geons, 
the production of geons will stop.
This is just like a structure formation by Jeans instability.

Nonetheless, extended analysis may be required to conclude the  nonlinear 
(in)stability of the Minkowski spacetime with massive gravitons.
We have also to study a quantum mechanical (in)stability
against quantum fluctuations.

Since the ground state of the massive graviton geon would be the quadrupole geon, the produced geons should eventually transit to the quadrupole geon. 
Although the resultant quadrupole geon will emit the gravitational waves, the gravitational wave emission is drastically suppressed if the produced geon is non-relativistic. As a result, we may expect that the geon may not  decay soon and the localized structure would remain.


Although we have only discussed the classical features of the Poisson-Schr\"{o}dinger equations \eqref{Poisson_eq} and \eqref{Schrodinger_eq}, one may quantize the massive graviton regarding $\psi_{ij}$ as indeed the wavefunction of the massive graviton. The ``Hamiltonian operator'' could be defined by
\begin{align}
\hat{H}:=-\frac{\Delta}{2m_G}+m_G \Phi\,.
\end{align}
The massive graviton geon may be interpreted as a Bose-Einstein condensate of the massive gravitons just by the analogy to the massive scalar field and then it can be seen as a macroscopic object. If the massive graviton geon can transit quantum mechanically from one state to another, this describes 
a quantum transition of spacetime since the spacetime metric is given by \eqref{g_metric} (or \eqref{eff_metric}). On the other hand, introducing a matter field and supposing the gravitational potential is dominated by the matter with a mass $M_{\star}$, i.e., $\Phi=-GM_{\star}/r$, the Schr\"{o}dinger equation is exactly solved in a way similar to the quantization of the Hydrogen atom. The non-relativistic massive graviton may be quantized by the method of the canonical quantization.

Finally, we notice that interesting phenomena must exist in relativistic extensions of our study. For instance, we need to discuss the relativistic system to obtain the precise value of the maximum mass of the geon. In the Newtonian limit, the quadrupole geon is degenerated because the projection angular momentum $j_z$ does not contribute to the equations of the geons. However, in the relativistic system, it 
is no longer true. We may expect that the degeneracy is resolved. Furthermore, the Vainshtein mechanism is irrelevant to the non-relativistic geons while it should be important for the relativistic geons. All these investigations require treatments beyond the Newtonian approximation.


\section*{Acknowledgments}
KA would like to thank the Institute of  Cosmology and Gravitation, University of Portsmouth for their hospitality during his visit.
KA also acknowledges Kazuya Koyama and Jiro Soda for useful discussions.
KM would like to thank Gary Gibbons for the information 
about the geons and the periodic solutions. 
This work was supported in part by  JSPS KAKENHI Grant Numbers JP15J05540 (KA), JP16K05362 (KM), JP17H06359 (KM), and JP17K18792 (HO).


\appendix

\section{Pure-spin and pure-orbital spherical harmonics}
\label{app_harmonics}

As is well-known, the angular dependence of a quantity (which can be either a scaler, a vector, or a tensor) can be expanded in terms of the spherical harmonics. Here, we focus on a symmetric and traceless three-dimensional tensor $t_{ij}$ and consider a multipole expansion of it. We introduce two types of the orthonormal bases of the multipole expansion which are called the pure-spin spherical harmonics $(Y^{A}_{j,j_z})_{ij}$ and the pure-orbital spherical harmonics $(T^{s}_{j,j_z})_{ij}$, respectively\footnote{If $t_{ij}$ is not traceless, the trace part can be expanded in terms of $Y_{j,j_z}\gamma_{ij}/\sqrt{3}$ which corresponds to the spin-zero state, i.e., $\hat{\mathbf{S}}^2 (Y_{j,j_z}\gamma_{ij})=0$.} (see \cite{Thorne:1980ru,Maggiore:1900zz} for more details). Note that the suffix $j$ inside the parenthesis represents the total angular momentum while the indices outside the parenthesis are the spatial indices of the tensor.

We first introduce the orbital angular momentum operators $\hat{\mathbf{L}}_I$ and the total angular momentum operators $\hat{\mathbf{J}}_I$ which satisfy the commutator relations
\begin{align}
[\hat{\mathbf{L}}_I,\hat{\mathbf{L}}_J]&=i\sum_K \epsilon_{IJK} \hat{\mathbf{L}}_K\,, \\
[\hat{\mathbf{J}}_I,\hat{\mathbf{J}}_J]&=i\sum_K \epsilon_{IJK} \hat{\mathbf{J}}_K\,, \\
[\hat{\mathbf{J}}_I,\hat{\mathbf{L}}_J]&=i\sum_K \epsilon_{IJK} \hat{\mathbf{L}}_K\,, 
\end{align}
where $I,J,K$ are the indices of the the Cartesian coordinates, $I,J,K=(x,y,z)$.
The operators are explicitly given by
\begin{align}
\hat{\mathbf{L}}_I:\!&=-i(\mathbf{r}\times \partial/\partial \mathbf{r} )_I \,, \nn
             &=-i\xi^i_I \partial_i \,,\\
\hat{\mathbf{J}}_I:\!&=-i\mathcal{L}_{\xi_{I}} \,,
\end{align} 
where $\xi^i_I$ is the rotational Killing vectors around $I=x,y,z$ axes and $\mathcal{L}_{\xi_{I}}$ are the Lie derivatives with respect to them. We recall that the indices $i,j,k$ are not necessary to be those of the the Cartesian coordinates. For instance, the Killing vectors are written by
\begin{align}
\xi_x^i&=(0,-\sin \phi, -\cot \theta \cos \phi) \,, \\
\xi_y^i&=(0,\cos \phi, -\cot \theta \sin \phi) \,, \\
\xi_z^i&=(0,0,1)\,,
\end{align}
in the spherical coordinates, $i,j,k=(r,\theta,\phi)$.
The spin angular momentum operators $\hat{\mathbf{S}}_I$ are defined so that $\hat{\mathbf{J}}_I=\hat{\mathbf{L}}_I+\hat{\mathbf{S}}_I$. The magnitude squared of the operators are defined by the sum of the squared operators, e.g.,
\begin{align}
\hat{\mathbf{L}}^2=\hat{\mathbf{L}}_x^2+\hat{\mathbf{L}}_y^2+\hat{\mathbf{L}}_z^2\,.
\end{align} 
We note 
\begin{align}
\hat{\mathbf{S}}^2t_{ij}=2\times (2+1) t_{ij}\,,
\end{align}
which explicitly shows that the symmetric and traceless tensor is a spin-2 field.

We assume that some tensor spherical harmonics $(Y_{j,j_z})_{ij}$ are the eigenstates of the operators $\hat{\mathbf{J}}^2$ and $\hat{\mathbf{J}}_z$:
\begin{align}
\hat{\mathbf{J}}^2(Y_{j,j_z})_{ij}&=j(j+1)(Y_{j,j_z})_{ij}\,, \\
\hat{\mathbf{J}}_z(Y_{j,j_z})_{ij}&=j_z (Y_{j,j_z})_{ij}\,,
\end{align}
where $j=0,1,2,\cdots$ and $j_z=0,\pm 1, \cdots, \pm j$.
However, these conditions do not determine the tensor spherical harmonics completely. We can further assume a property of them.

The pure-spin spherical harmonics $(Y^A_{j,j_z})_{ij}$ are understood as polarization eigenstates of a spherical wave of a spin-2 field where the polarization states are labeled by $A=S_0,E_1,E_2,B_1,B_2$. The propagating direction of the spherical wave is given by the unit radial vector $n^i=r^i/|r|$. The spin-2 polarization modes $E_2,B_2$ are transverse to the propagating direction,
\begin{align}
n^i (Y^{E_2}_{j,j_z})_{ij}=n^i (Y^{B_2}_{j,j_z})_{ij}&=0 \,, 
\end{align}
the spin-1 polarization modes $E_1,B_1$ are mixed longitudinal and transverse
\begin{align}
n^i (Y^{E_1}_{j,j_z})_{ij}\neq 0\,, \quad  n^i (Y^{B_1}_{j,j_z})_{ij}&\neq 0 \,, \nn
n^i n^j (Y^{E_1}_{j,j_z})_{ij}=n^i n^j (Y^{B_1}_{j,j_z})_{ij}&=0 \,,
\end{align}
and the spin-0 polarization mode $S_0$ is the longitudinal mode
\begin{align}
n^i (Y^{S_0}_{j,j_z})_{ij}\neq 0\,, \quad  n^i n^j (Y^{S_0}_{j,j_z})_{ij}&\neq 0 \,.
\end{align}
The modes $E_1$ and $E_2$ are called the electric-type parity, i.e., the even parity according to $(\theta,\phi) \rightarrow (\pi-\theta,\pi+\phi)$ while $B_1$ and $B_2$ are the magnetic-type parity, i.e., the odd parity. The pure-spin spherical harmonics are orthonormal
\begin{align}
\int d \Omega (Y^{A}_{j,j_z} )_{ij}^* (Y^{A'}_{j',j_z'} ){}^{ij}&=\delta_{AA'}\delta_{jj'}\delta_{j_z j_z'} \,,
\end{align}
where the complex conjugate is given by $(Y^{A}_{j,j_z} )_{ij}^*=(-1)^{j_z} (Y^{A}_{j,-j_z} )_{ij}$. In the spherical coordinates, they are explicitly given by
\begin{align}
(Y^{S_0}_{j,j_z})_{ij}&=\frac{1}{\sqrt{6}}
\begin{pmatrix}
2Y_{j,j_z}& 0 \\
*& -r^2 Y_{j,j_z} \hat{\gamma}_{ab}
\end{pmatrix}
\,,\nn
(Y^{E_1}_{j,j_z})_{ij}&=
\begin{pmatrix}
0 & r (Y_{j,j_z})_b\\
* & 0
\end{pmatrix}
\,, \nn
(Y^{E_2}_{j,j_z})_{ij}&=
\begin{pmatrix}
0 & 0 \\
* & r^2 (Y_{j,j_z})_{ab} 
\end{pmatrix}
\,, \nn
(Y^{B_1}_{j,j_z})_{ij}&=
\begin{pmatrix}
0 & r \epsilon_b{}^c(Y_{j,j_z})_c\\
* & 0
\end{pmatrix}
\,, \nn
(Y^{B_2}_{j,j_z})_{ij}&=
\begin{pmatrix}
0 & 0 \\
* & r^2 \epsilon_{(a}{}^c(Y_{j,j_z})_{b)c} 
\end{pmatrix}
\,, \label{pure_spin}
\end{align}
with
\begin{align}
(Y_{j,j_z})_a &=-\frac{1}{\sqrt{2j(j+1)}}\hat{D}_a Y_{j,j_z}\,, \\
(Y_{j,j_z})_{ab} &=\sqrt{\frac{2}{(j-1)j(j+1)(j+2)}}
\nn& \qquad \times \left(\hat{D}_a \hat{D}_b-\frac{1}{2}\hat{D}^2 \hat{\gamma}_{ab} \right)Y_{j,j_z}
\,, 
\end{align}
and the spherical harmonics $Y_{j,j_z}$ where $a,b,c$ are the indices in the unit two-sphere $d^2\Omega=\hat{\gamma}_{ab}dx^a dx^b=d^2\theta+\sin^2 \theta d \phi^2$ and $\hat{D}_a$ is the covariant derivative with respect to $\hat{\gamma}_{ab}$. The two-dimensional Levi-Civita tensor is denoted by $\epsilon_{ab}$. Clearly from the explicit forms, the modes $E_1,B_1$ are defined only for $j \geq 1$ and the modes $E_2,B_2$ are defined only for $j \geq 2$. In Table \ref{table_def}, we summarize which tensor spherical harmonics are defined in each total angular momentum.

\begin{table}[tb]
\caption{Defined harmonics in each total angular momentum.}
\label{table_def}
\begin{tabular}{ccccc}
\hline
            & pure-spin & pure-orbital \\ 
            & (helicity) & (orbital angular momentum) \\
\hline\hline
$j=0$     & $A=S_0$                 & $s=-2$ \\
          & $(h=0)$                 & $(\ell=2)$ \\
$\,j=1\,$ & $\,\,A=S_0,E_1,B_1\,\,$ & $s=0,-1,-2$ \\
          & $(h=0,\pm 1)$           & $(\ell=1,2,3)$ \\
$j\geq2$  & \multicolumn{2}{c}{all modes are defined} \\
          & $(h=0, \pm 1, \pm 2)$   & $(\ell=0,1,2,\cdots)$ \\
\hline
\end{tabular}
\end{table}

The helicity eigenstates $(Y^h_{j,j_z})_{ij}$ are defined by
\begin{align}
\mathbf{n}\cdot \hat{\mathbf{S}} (Y^h_{j,j_z})_{ij}=h(Y^h_{j,j_z})_{ij} \,,
\end{align}
where $\mathbf{n}$ is the unit radial vector in the Cartesian coordinates and $h=0, \pm 1, \pm 2$. The helicity eigenstates are given by linear combinations of the pure-spin spherical harmonics as
\begin{align}
(Y^0_{j,j_z})_{ij}&=(Y^{S_0}_{j,j_z})_{ij} \,, \nn
(Y^{+1}_{j,j_z})_{ij}&=\frac{1}{\sqrt{2}}\left[ (Y^{E_1}_{j,j_z})_{ij}+i (Y^{B_1}_{j,j_z})_{ij} \right] \,, \nn
(Y^{-1}_{j,j_z})_{ij}&=\frac{1}{\sqrt{2}}\left[ (Y^{E_1}_{j,j_z})_{ij}-i (Y^{B_1}_{j,j_z})_{ij} \right] \,, \nn
(Y^{+2}_{j,j_z})_{ij}&=\frac{1}{\sqrt{2}}\left[ (Y^{E_2}_{j,j_z})_{ij}+i (Y^{B_2}_{j,j_z})_{ij} \right] \,, \nn
(Y^{-2}_{j,j_z})_{ij}&=\frac{1}{\sqrt{2}}\left[ (Y^{E_2}_{j,j_z})_{ij}-i (Y^{B_2}_{j,j_z})_{ij} \right] \,,
\end{align}
which also indicates that $A=S_0,$ $A=E_1,B_1,$ and $A=E_2,B_2$ represent the polarization states of the spin-2 field, respectively. In the $j=0$ mode, there exists only the helicity-zero mode and there are the $h=0,\pm 1$ modes in the $j=1$ modes. The helicity-two modes appear when $j\geq 2$. In a massless spin-2 field, the helicity-zero mode and the helicity-one modes are not dynamical degrees of freedom whereas all modes are dynamical in the massive case.

On the other hand, the pure-orbital spherical harmonics $(T^s_{j,j_z})_{ij}$ are defined by the eigenstates of the orbital angular momentum operator
\begin{align}
\hat{\mathbf{L}}^2 (T^{s}_{j,j_z})_{ij} &= \ell (\ell+1) (T^{s}_{j,j_z})_{ij} \,, 
\end{align}
with the orthonormality
\begin{align}
\int d \Omega (T^{s}_{j,j_z} )_{ij}^* (T^{s'}_{j',j_z'} ){}^{ij}&=\delta_{ss'}\delta_{jj'}\delta_{j_z j_z'} \,.
\end{align}
The possible total angular momenta of the spin-2 field are given by $j=\ell,\ell \pm 1, \ell \pm 2$ which are labeled by $s=0,\pm 1,\pm 2$, i.e., $j=\ell+s$. We note that the label $s$ does not represent the spin states since the pure-orbital spherical harmonics is no longer the eigenstates of the polarizations of the spin-2 field. The pure-orbital spherical harmonics are characterized by their angular momenta. We note that
\begin{align}
\hat{\mathbf{L}}^2=-r^2 \Delta +r^ir^j\partial_i \partial_j+2r^i \partial_i \,,
\end{align}
thus, the pure-orbital spherical harmonics are also eigenstates of the Laplace operator. As a result, we obtain
\begin{align}
\Delta \left[ f(r) (T^s_{j,j_z})_{ij}\right]&=\left[\frac{1}{r^2}\frac{d}{dr}r^2 \frac{d}{dr}f-\frac{\ell(\ell+1)}{r^2}f\right](T^{s}_{j,j_z})_{ij}\,,
\end{align}
where $f$ is a function of $r$.
The pure-orbital spherical harmonics are explicitly constructed by the so-called spin function as well as the (scalar) spherical harmonics (see \cite{Thorne:1980ru,Maggiore:1900zz}). 
Here, we only show the relations between the pure-spin spherical harmonics and the pure-orbital spherical harmonics by which one can explicitly obtain the pure-orbital spherical harmonics from Eq.~\eqref{pure_spin}. The pure-orbital spherical harmonics are obtained by the orthogonal transformation as
\begin{align}
(T^{-2}_{j,j_z})_{ij} &=\sqrt{\frac{3(j+1)(j+2)}{2(2j+1)(2j+3)}}(Y^{S_0}_{j,j_z})_{ij}
\nn&
+\sqrt{\frac{2j(j+2)}{(2j+1)(2j+3)}}(Y^{E_1}_{j,j_z})_{ij}
\nn 
&+\sqrt{\frac{j(j-1)}{2(2j+1)(2j+3)}} (Y^{E_2}_{j,j_z})_{ij} \,,
\nn
(T^{0}_{j,j_z})_{ij} &=-\sqrt{\frac{j(j +1)}{(2j-1)(2j+3)}} (Y^{S_0}_{j,j_z})_{ij}
\nn&
+\sqrt{\frac{3}{(2j-1)(2j+3)}}(Y^{E_1}_{j,j_z})_{ij}
\nn 
&+\sqrt{\frac{3(j-1)(j+2)}{(2j-1)(2j+3)}}(Y^{E_2}_{j,j_z})_{ij}
\,, \nn
(T^{+2}_{j,j_z})_{ij} &=
 \sqrt{\frac{3j(j-1)}{2(2j+1)(2j-1)}} (Y^{S_0}_{j,j_z})_{ij}
\nn&
-\sqrt{\frac{2(j-1)(j+1)}{(2j+1)(2j-1)}} (Y^{E_1}_{j,j_z})_{ij}
\nn
&+\sqrt{\frac{(j+1)(j+2)}{2(2j+1)(2j-1)}} (Y^{E_2}_{j,j_z})_{ij}
\,, \nn
(T^{-1}_{j,j_z})_{ij}&=\sqrt{\frac{j+2}{2j+1}} (Y^{B_1}_{j,j_z})_{ij} + \sqrt{\frac{j-1}{2j+1}} (Y^{B_2}_{j,j_z})_{ij} \,, \nn
(T^{+1}_{j,j_z})_{ij}&=\sqrt{\frac{j-1}{2j+1}} (Y^{B_1}_{j,j_z})_{ij} -\sqrt{\frac{j+2}{2j+1}} ( Y^{B_2}_{j,j_z})_{ij} \,. \label{ortho}
\end{align}
We note that $(T^{0}_{j,j_z})$ and $(T^{-1}_{j,j_z})$ are defined only for $j\geq 1$ whereas $(T^{+2}_{j,j_z})$ and $(T^{+1}_{j,j_z})$ are defined only for $j\geq 2$ (see Table \ref{table_def}). The relations \eqref{ortho} can be used even for $j=0,1$ since the coefficients in front of the undefined variables in the right-hand side vanish when $j=0,1$. 

The total angular momentum operators are given by the Lie derivatives. The symmetry of the field configuration is determined by the total angular momentum. For instance, the spherically symmetric mode is $j=j_z=0$ and the axisymmetric modes are $j_z=0$. On the other hand, the orbital angular momentum vanishes only for the $j=s=2$ modes of the pure-orbital spherical harmonics which are not spherically symmetric. 

We find that the $j=s=2$ pure-orbital spherical harmonics further have a special property: 
orthonormal \emph{without} integration
\begin{align}
(T^{+2}_{2, j_z} )_{ij}^* (Y^{+2}_{2, j_z'} ){}^{ij}=\frac{1}{4\pi}\delta_{j_z j_z'}\,,
\end{align}
which hold only for the $j=2$ modes of $(T^{+2}_{j, j_z} )_{ij}$. The explicit forms in the spherical coordinates are given by
\begin{align}
&(T^{+2}_{2,0})_{ij}=\frac{1}{\sqrt{16\pi}}
\nn
&\times 
\begin{pmatrix}
\frac{1}{\sqrt{6}}(1+3\cos 2\theta) & -\frac{\sqrt{6}}{2}r \sin 2\theta & 0 \\
* & \frac{r^2}{\sqrt{6}}(1-3 \cos 2\theta) & 0 \\
* & * &-\sqrt{\frac{2}{3}}r^2 \sin^2 \theta
\end{pmatrix}
, \\
&(T^{+2}_{2,1})_{ij}=\frac{e^{i\phi}}{\sqrt{16\pi}}
\begin{pmatrix}
-\sin 2 \theta & -r \cos 2\theta & -i r \sin \theta \cos \theta  \\
* & r^2 \sin 2 \theta & i r^2 \sin^2\theta \\
* & * & 0
\end{pmatrix}
, \\
&(T^{+2}_{2,2})_{ij}=\frac{e^{2i \phi}}{\sqrt{16\pi}}
\begin{pmatrix}
\sin^2 \theta & r \sin \theta \cos \theta & ir \sin^2 \theta \\
* & r^2 \cos^2 \theta & i r^2 \sin \theta \cos \theta \\
* & * & -r^2 \sin^2 \theta
\end{pmatrix}
,
\end{align}
and $j_z<0$ are obtained by the relation $(T^{+2}_{2,-j_z})_{ij}=(-1)^{j_z} (T^{2}_{2,j_z})^*_{ij}$. On the other hand, in the Cartesian coordinates, they are just constant matrices
\begin{align}
(T^{+2}_{2,0})_{ij} &=\frac{1}{\sqrt{24\pi}}
\begin{pmatrix}
-1 & 0 & 0 \\
* & -1 & 0 \\
* & *  & 2 
\end{pmatrix}
\,, \\
(T^{+2}_{2,\pm 1})_{ij} &=\mp \frac{1}{\sqrt{16\pi}}
\begin{pmatrix}
0 & 0 & 1 \\
* & 0 & \pm i \\
* & *  & 0 
\end{pmatrix}
\,, \\
(T^{+2}_{2,\pm 2})_{ij} &= \frac{1}{\sqrt{16\pi}}
\begin{pmatrix}
1 & \pm i & 0 \\
* & -1 & 0 \\
* & *  & 0 
\end{pmatrix}
\,.
\end{align}

\section{Octupole geons}
\label{sec_oct}

\begin{table}[tb]
\caption{Energy eigenvalues and $95\%$-mass radii of octupole geon.}
\label{table_ell=3}
\begin{tabular}{ccccc}
\hline
$\quad $   node     $\quad $      & $\tilde{E}$ & $\quad  \tilde{R}_{95\%} \quad $ \\
\hline\hline
$n=0$ & $-0.05411 $ & $20.30$ \\
$n=1$ & $-0.01760 $ & $60.23$ \\
$n=2$ & $-0.008631$ & $120.6$ \\
$n=3$ & $-0.005109$ & $201.4$ \\
\hline
\end{tabular}
\end{table}

An octupole configuration also gives a spherically symmetric energy density distribution. We find
\begin{align}
(T^{+2}_{3, 2} )_{ij}^* (T^{+2}_{3, 2} ){}^{ij}=(T^{+2}_{3, -2} )_{ij}^* (T^{+2}_{3, -2} ){}^{ij}&=\frac{1}{4\pi} \,, \\
  (T^{+2}_{3, 2} )_{ij}^* (T^{+2}_{3, -2} ){}^{ij}&=0\,,
\end{align}
where
\begin{align}
&(T^{+2}_{3, 2} )_{ij}=\frac{e^{2i\phi}}{\sqrt{16\pi}} \times \nn
&\begin{pmatrix}
3 \cos \theta \sin^2 \theta & r(1+3\cos2\theta)\sin \theta/2 & 2 ir \cos \theta \sin^2 \theta \\
* & r^2(\cos \theta+3\cos 3 \theta)/4 & ir^2 \cos 2\theta \sin \theta \\
* & * & -r^2 \cos \theta \sin^2 \theta 
\end{pmatrix}
.
\end{align}
Hence, if the configuration of the massive graviton is given by
\begin{align}
\psi_{ij}=\sqrt{16\pi}e^{-iEt}\psi_3(r)\left[a_+(T^{+2}_{3, 2} )_{ij}+a_- (T^{+2}_{3,-2} )_{ij} \right]\,,
\end{align}
we obtain $\Phi=\Phi(r)$ where $a_+$ and $a_-$ are constants with $a_+^2+a_-^2=1$. The orbital angular momentum of this mode is $\ell=1$.

The solutions can be found under the boundary conditions $\psi_3 \rightarrow 0$ both at the center and at infinity. We note $d\psi_3/dr$ does not vanish at the center because the $j=3$ modes are not symmetric about the equatorial plane. On the other hand, the boundary conditions of the gravitational potential are $d\Phi/dr \rightarrow 0$ at the center and $\Phi \rightarrow 0$ at infinity.

A few lower energy eigenvalues and the $95\%$-mass radii are summarized in Table \ref{table_ell=3}. Since the orbital angular momentum is smaller than the monopole geon $(\ell=2)$ but larger than the quadrupole geon $(\ell=0)$, the energy eigenvalues are intermediate between them.

Although the energy eigenvalue of the octupole geon is smaller than that of the monopole geon, we could not find the unstable mode in the linear stability analysis of the $j=3$ mode perturbations in Sec.~\ref{sec_perturbation}. This is not surprising because the linear stability does not conclude the background is indeed stable. No unstable mode in the $j=3$ mode linear perturbations implies that the monopole geon does not immediately transit to the octupole geon even if the transition is possible due to the nonlinear effects.

\bibliography{ref}

\providecommand{\href}[2]{#2}\begingroup\raggedright\begin{thebibliography}{10}

\bibitem{Abbott:2016blz}
{\scshape Virgo, LIGO Scientific} collaboration, B.~P. Abbott et~al.,
  \emph{{Observation of Gravitational Waves from a Binary Black Hole Merger}},
  \href{http://dx.doi.org/10.1103/PhysRevLett.116.061102}{\emph{Phys. Rev.
  Lett.} {\bf 116} (2016) 061102}, [\href{http://arxiv.org/abs/1602.03837}{{\tt
  1602.03837}}].

\bibitem{TheLIGOScientific:2016pea}
{\scshape Virgo, LIGO Scientific} collaboration, B.~P. Abbott et~al.,
  \emph{{Binary Black Hole Mergers in the first Advanced LIGO Observing Run}},
  \href{http://dx.doi.org/10.1103/PhysRevX.6.041015}{\emph{Phys. Rev.} {\bf X6}
  (2016) 041015}, [\href{http://arxiv.org/abs/1606.04856}{{\tt 1606.04856}}].

\bibitem{Abbott:2017vtc}
{\scshape VIRGO, LIGO Scientific} collaboration, B.~P. Abbott et~al.,
  \emph{{GW170104: Observation of a 50-Solar-Mass Binary Black Hole Coalescence
  at Redshift 0.2}},
  \href{http://dx.doi.org/10.1103/PhysRevLett.118.221101}{\emph{Phys. Rev.
  Lett.} {\bf 118} (2017) 221101}, [\href{http://arxiv.org/abs/1706.01812}{{\tt
  1706.01812}}].

\bibitem{Abbott:2017oio}
{\scshape Virgo, LIGO Scientific} collaboration, B.~P. Abbott et~al.,
  \emph{{GW170814: A Three-Detector Observation of Gravitational Waves from a
  Binary Black Hole Coalescence}},
  \href{http://dx.doi.org/10.1103/PhysRevLett.119.141101}{\emph{Phys. Rev.
  Lett.} {\bf 119} (2017) 141101}, [\href{http://arxiv.org/abs/1709.09660}{{\tt
  1709.09660}}].

\bibitem{Wheeler:1955zz}
J.~A. Wheeler, \emph{{Geons}},
  \href{http://dx.doi.org/10.1103/PhysRev.97.511}{\emph{Phys. Rev.} {\bf 97}
  (1955) 511--536}.

\bibitem{Brill:1964zz}
D.~R. Brill and J.~B. Hartle, \emph{{Method of the Self-Consistent Field in
  General Relativity and its Application to the Gravitational Geon}},
  \href{http://dx.doi.org/10.1103/PhysRev.135.B271}{\emph{Phys. Rev.} {\bf 135}
  (1964) B271--B278}.

\bibitem{Anderson:1996pu}
P.~R. Anderson and D.~R. Brill, \emph{{Gravitational geons revisited}},
  \href{http://dx.doi.org/10.1103/PhysRevD.56.4824}{\emph{Phys. Rev.} {\bf D56}
  (1997) 4824--4833}, [\href{http://arxiv.org/abs/gr-qc/9610074}{{\tt
  gr-qc/9610074}}].

\bibitem{Trautman:1957zz}
A.~Trautman, \emph{{Proof of the non-existence of periodic gravitational fields
  representing radiation}}, {\emph{Bull. Acad. Pol. Sci. Cl.3} {\bf 5} (1957)
  1115}.

\bibitem{Detweiler:1993eq}
S.~L. Detweiler, \emph{{Periodic solutions of the Einstein equations}},
  \href{http://dx.doi.org/10.1103/PhysRevD.50.4929}{\emph{Phys. Rev.} {\bf D50}
  (1994) 4929--4943}, [\href{http://arxiv.org/abs/gr-qc/9312016}{{\tt
  gr-qc/9312016}}].

\bibitem{gibbons1984absence}
G.~Gibbons and J.~Stewart, \emph{Absence of asymptotically flat solutions of
  einstein's equations which are periodic and empty near infinity},  in
  \emph{Classical general relativity. Proceedings of the conference on
  classical (non-quantum) general relativity, City University, London, 21-22
  December 1983}, 1984.

\bibitem{Tod:2009em}
P.~Tod, \emph{{On analytic asymptotically-flat vacuum and electrovac metrics,
  periodic in time}},  \href{http://arxiv.org/abs/0902.1061}{{\tt 0902.1061}}.

\bibitem{Bicak:2010xp}
J.~Bicak, M.~Scholtz and P.~Tod, \emph{{On asymptotically flat solutions of
  Einstein's equations periodic in time I. Vacuum and electrovacuum
  solutions}},
  \href{http://dx.doi.org/10.1088/0264-9381/27/5/055007}{\emph{Class. Quant.
  Grav.} {\bf 27} (2010) 055007}, [\href{http://arxiv.org/abs/1003.3402}{{\tt
  1003.3402}}].

\bibitem{Bicak:2010tt}
J.~Bicak, M.~Scholtz and P.~Tod, \emph{{On asymptotically flat solutions of
  Einstein's equations periodic in time II. Spacetimes with scalar-field
  sources}},
  \href{http://dx.doi.org/10.1088/0264-9381/27/17/175011}{\emph{Class. Quant.
  Grav.} {\bf 27} (2010) 175011}, [\href{http://arxiv.org/abs/1008.0248}{{\tt
  1008.0248}}].

\bibitem{Alexakis:2015ara}
S.~Alexakis and V.~Schlue, \emph{{Non-existence of time-periodic vacuum
  spacetimes}},  \href{http://arxiv.org/abs/1504.04592}{{\tt 1504.04592}}.

\bibitem{Enriquez:2016klf}
R.~T. Enriquez, \emph{{Stationarity of asymptotically flat non-radiating
  electrovacuum spacetimes}},  \href{http://arxiv.org/abs/1607.04882}{{\tt
  1607.04882}}.

\bibitem{Dias:2012tq}
O.~J.~C. Dias, G.~T. Horowitz, D.~Marolf and J.~E. Santos, \emph{{On the
  Nonlinear Stability of Asymptotically Anti-de Sitter Solutions}},
  \href{http://dx.doi.org/10.1088/0264-9381/29/23/235019}{\emph{Class. Quant.
  Grav.} {\bf 29} (2012) 235019}, [\href{http://arxiv.org/abs/1208.5772}{{\tt
  1208.5772}}].

\bibitem{Horowitz:2014hja}
G.~T. Horowitz and J.~E. Santos, \emph{{Geons and the Instability of Anti-de
  Sitter Spacetime}},
  \href{http://dx.doi.org/10.4310/SDG.2015.v20.n1.a13}{\emph{Surveys Diff.
  Geom.} {\bf 20} (2015) 321--335}, [\href{http://arxiv.org/abs/1408.5906}{{\tt
  1408.5906}}].

\bibitem{Martinon:2017uyo}
G.~Martinon, G.~Fodor, P.~Grandcl{\'e}ment and P.~Forg{\'a}cs,
  \emph{{Gravitational geons in asymptotically anti-de Sitter spacetimes}},
  \href{http://dx.doi.org/10.1088/1361-6382/aa6f48}{\emph{Class. Quant. Grav.}
  {\bf 34} (2017) 125012}, [\href{http://arxiv.org/abs/1701.09100}{{\tt
  1701.09100}}].

\bibitem{deRham:2014zqa}
C.~de~Rham, \emph{{Massive Gravity}},
  \href{http://dx.doi.org/10.12942/lrr-2014-7}{\emph{Living Rev. Rel.} {\bf 17}
  (2014) 7}, [\href{http://arxiv.org/abs/1401.4173}{{\tt 1401.4173}}].

\bibitem{Schmidt-May:2015vnx}
A.~Schmidt-May and M.~von Strauss, \emph{{Recent developments in bimetric
  theory}}, \href{http://dx.doi.org/10.1088/1751-8113/49/18/183001}{\emph{J.
  Phys.} {\bf A49} (2016) 183001}, [\href{http://arxiv.org/abs/1512.00021}{{\tt
  1512.00021}}].

\bibitem{Feng:2003nr}
J.~L. Feng, A.~Rajaraman and F.~Takayama, \emph{{Graviton cosmology in
  universal extra dimensions}},
  \href{http://dx.doi.org/10.1103/PhysRevD.68.085018}{\emph{Phys. Rev.} {\bf
  D68} (2003) 085018}, [\href{http://arxiv.org/abs/hep-ph/0307375}{{\tt
  hep-ph/0307375}}].

\bibitem{Dubovsky:2004ud}
S.~L. Dubovsky, P.~G. Tinyakov and I.~I. Tkachev, \emph{{Massive graviton as a
  testable cold dark matter candidate}},
  \href{http://dx.doi.org/10.1103/PhysRevLett.94.181102}{\emph{Phys. Rev.
  Lett.} {\bf 94} (2005) 181102},
  [\href{http://arxiv.org/abs/hep-th/0411158}{{\tt hep-th/0411158}}].

\bibitem{Pshirkov:2008nr}
M.~Pshirkov, A.~Tuntsov and K.~A. Postnov, \emph{{Constraints on the massive
  graviton dark matter from pulsar timing and precision astrometry}},
  \href{http://dx.doi.org/10.1103/PhysRevLett.101.261101}{\emph{Phys. Rev.
  Lett.} {\bf 101} (2008) 261101}, [\href{http://arxiv.org/abs/0805.1519}{{\tt
  0805.1519}}].

\bibitem{Aoki:2016zgp}
K.~Aoki and S.~Mukohyama, \emph{{Massive gravitons as dark matter and
  gravitational waves}},
  \href{http://dx.doi.org/10.1103/PhysRevD.94.024001}{\emph{Phys. Rev.} {\bf
  D94} (2016) 024001}, [\href{http://arxiv.org/abs/1604.06704}{{\tt
  1604.06704}}].

\bibitem{Babichev:2016hir}
E.~Babichev, L.~Marzola, M.~Raidal, A.~Schmidt-May, F.~Urban, H.~Veerm{\"a}e
  et~al., \emph{{Bigravitational origin of dark matter}},
  \href{http://dx.doi.org/10.1103/PhysRevD.94.084055}{\emph{Phys. Rev.} {\bf
  D94} (2016) 084055}, [\href{http://arxiv.org/abs/1604.08564}{{\tt
  1604.08564}}].

\bibitem{Babichev:2016bxi}
E.~Babichev, L.~Marzola, M.~Raidal, A.~Schmidt-May, F.~Urban, H.~Veerm{\"a}e
  et~al., \emph{{Heavy spin-2 Dark Matter}},
  \href{http://dx.doi.org/10.1088/1475-7516/2016/09/016}{\emph{JCAP} {\bf 1609}
  (2016) 016}, [\href{http://arxiv.org/abs/1607.03497}{{\tt 1607.03497}}].

\bibitem{Aoki:2017cnz}
K.~Aoki and K.-i. Maeda, \emph{{Condensate of Massive Graviton and Dark
  Matter}},  \href{http://arxiv.org/abs/1707.05003}{{\tt 1707.05003}}.

\bibitem{Aoki:2017ffl}
K.~Aoki and S.~Mukohyama, \emph{{Massive graviton dark matter with environment
  dependent mass: A natural explanation of the dark matter-baryon ratio}},
  \href{http://dx.doi.org/10.1103/PhysRevD.96.104039}{\emph{Phys. Rev.} {\bf
  D96} (2017) 104039}, [\href{http://arxiv.org/abs/1708.01969}{{\tt
  1708.01969}}].

\bibitem{Marzola:2017lbt}
L.~Marzola, M.~Raidal and F.~R. Urban, \emph{{Oscillating Spin-2 Dark Matter}},
   \href{http://arxiv.org/abs/1708.04253}{{\tt 1708.04253}}.

\bibitem{Chu:2017msm}
X.~Chu and C.~Garcia-Cely, \emph{{Self-interacting Spin-2 Dark Matter}},
  \href{http://dx.doi.org/10.1103/PhysRevD.96.103519}{\emph{Phys. Rev.} {\bf
  D96} (2017) 103519}, [\href{http://arxiv.org/abs/1708.06764}{{\tt
  1708.06764}}].

\bibitem{Albornoz:2017yup}
N.~L.~G. Albornoz, A.~Schmidt-May and M.~von Strauss, \emph{{Dark matter
  scenarios with multiple spin-2 fields}},
  \href{http://arxiv.org/abs/1709.05128}{{\tt 1709.05128}}.

\bibitem{Garny:2017kha}
M.~Garny, A.~Palessandro, M.~Sandora and M.~S. Sloth, \emph{{Theory and
  Phenomenology of Planckian Interacting Massive Particles as Dark Matter}},
  \href{http://arxiv.org/abs/1709.09688}{{\tt 1709.09688}}.

\bibitem{Seidel:1991zh}
E.~Seidel and W.~M. Suen, \emph{{Oscillating soliton stars}},
  \href{http://dx.doi.org/10.1103/PhysRevLett.66.1659}{\emph{Phys. Rev. Lett.}
  {\bf 66} (1991) 1659--1662}.

\bibitem{Kaup:1968zz}
D.~J. Kaup, \emph{{Klein-Gordon Geon}},
  \href{http://dx.doi.org/10.1103/PhysRev.172.1331}{\emph{Phys. Rev.} {\bf 172}
  (1968) 1331--1342}.

\bibitem{Jetzer:1991jr}
P.~Jetzer, \emph{{Boson stars}},
  \href{http://dx.doi.org/10.1016/0370-1573(92)90123-H}{\emph{Phys. Rept.} {\bf
  220} (1992) 163--227}.

\bibitem{Brito:2015pxa}
R.~Brito, V.~Cardoso, C.~A.~R. Herdeiro and E.~Radu, \emph{{Proca stars:
  Gravitating Bose-Einstein condensates of massive spin 1 particles}},
  \href{http://dx.doi.org/10.1016/j.physletb.2015.11.051}{\emph{Phys. Lett.}
  {\bf B752} (2016) 291--295}, [\href{http://arxiv.org/abs/1508.05395}{{\tt
  1508.05395}}].

\bibitem{Schunck:2003kk}
F.~E. Schunck and E.~W. Mielke, \emph{{General relativistic boson stars}},
  \href{http://dx.doi.org/10.1088/0264-9381/20/20/201}{\emph{Class. Quant.
  Grav.} {\bf 20} (2003) R301--R356},
  [\href{http://arxiv.org/abs/0801.0307}{{\tt 0801.0307}}].

\bibitem{Herdeiro:2017fhv}
C.~A.~R. Herdeiro, A.~M. Pombo and E.~Radu, \emph{{Asymptotically flat scalar,
  Dirac and Proca stars: discrete vs. continuous families of solutions}},
  \href{http://dx.doi.org/10.1016/j.physletb.2017.09.036}{\emph{Phys. Lett.}
  {\bf B773} (2017) 654--662}, [\href{http://arxiv.org/abs/1708.05674}{{\tt
  1708.05674}}].

\bibitem{Hassan:2011zd}
S.~F. Hassan and R.~A. Rosen, \emph{{Bimetric Gravity from Ghost-free Massive
  Gravity}}, \href{http://dx.doi.org/10.1007/JHEP02(2012)126}{\emph{JHEP} {\bf
  02} (2012) 126}, [\href{http://arxiv.org/abs/1109.3515}{{\tt 1109.3515}}].

\bibitem{Isaacson:1968zza}
R.~A. Isaacson, \emph{{Gravitational Radiation in the Limit of High Frequency.
  II. Nonlinear Terms and the Ef fective Stress Tensor}},
  \href{http://dx.doi.org/10.1103/PhysRev.166.1272}{\emph{Phys. Rev.} {\bf 166}
  (1968) 1272--1279}.

\bibitem{deRham:2010ik}
C.~de~Rham and G.~Gabadadze, \emph{{Generalization of the Fierz-Pauli Action}},
  \href{http://dx.doi.org/10.1103/PhysRevD.82.044020}{\emph{Phys. Rev.} {\bf
  D82} (2010) 044020}, [\href{http://arxiv.org/abs/1007.0443}{{\tt
  1007.0443}}].

\bibitem{deRham:2010kj}
C.~de~Rham, G.~Gabadadze and A.~J. Tolley, \emph{{Resummation of Massive
  Gravity}},
  \href{http://dx.doi.org/10.1103/PhysRevLett.106.231101}{\emph{Phys. Rev.
  Lett.} {\bf 106} (2011) 231101}, [\href{http://arxiv.org/abs/1011.1232}{{\tt
  1011.1232}}].

\bibitem{Vainshtein:1972sx}
A.~I. Vainshtein, \emph{{To the problem of nonvanishing gravitation mass}},
  \href{http://dx.doi.org/10.1016/0370-2693(72)90147-5}{\emph{Phys. Lett.} {\bf
  B39} (1972) 393--394}.

\bibitem{Ruffini:1969qy}
R.~Ruffini and S.~Bonazzola, \emph{{Systems of selfgravitating particles in
  general relativity and the concept of an equation of state}},
  \href{http://dx.doi.org/10.1103/PhysRev.187.1767}{\emph{Phys. Rev.} {\bf 187}
  (1969) 1767--1783}.

\bibitem{bernstein1998eigenstates}
D.~H. Bernstein, E.~Giladi and K.~R. Jones, \emph{Eigenstates of the
  gravitational schr{\"o}dinger equation}, {\emph{Modern Physics Letters A}
  {\bf 13} (1998) 2327--2336}.

\bibitem{Moroz:1998dh}
I.~M. Moroz, R.~Penrose and P.~Tod, \emph{{Spherically symmetric solutions of
  the Schrodinger-Newton equations}},
  \href{http://dx.doi.org/10.1088/0264-9381/15/9/019}{\emph{Class. Quant.
  Grav.} {\bf 15} (1998) 2733--2742}.

\bibitem{harrison2002numerical}
R.~Harrison, I.~Moroz and K.~Tod, \emph{A numerical study of the
  schr{\"o}dinger--newton equations}, {\emph{Nonlinearity} {\bf 16} (2002)
  101}.

\bibitem{Guzman:2004wj}
F.~S. Guzman and L.~A. Urena-Lopez, \emph{{Evolution of the Schrodinger-Newton
  system for a selfgravitating scalar field}},
  \href{http://dx.doi.org/10.1103/PhysRevD.69.124033}{\emph{Phys. Rev.} {\bf
  D69} (2004) 124033}, [\href{http://arxiv.org/abs/gr-qc/0404014}{{\tt
  gr-qc/0404014}}].

\bibitem{Hui:2016ltb}
L.~Hui, J.~P. Ostriker, S.~Tremaine and E.~Witten, \emph{{Ultralight scalars as
  cosmological dark matter}},
  \href{http://dx.doi.org/10.1103/PhysRevD.95.043541}{\emph{Phys. Rev.} {\bf
  D95} (2017) 043541}, [\href{http://arxiv.org/abs/1610.08297}{{\tt
  1610.08297}}].

\bibitem{Babichev:2013una}
E.~Babichev and A.~Fabbri, \emph{{Instability of black holes in massive
  gravity}},
  \href{http://dx.doi.org/10.1088/0264-9381/30/15/152001}{\emph{Class. Quant.
  Grav.} {\bf 30} (2013) 152001}, [\href{http://arxiv.org/abs/1304.5992}{{\tt
  1304.5992}}].

\bibitem{Brito:2013wya}
R.~Brito, V.~Cardoso and P.~Pani, \emph{{Massive spin-2 fields on black hole
  spacetimes: Instability of the Schwarzschild and Kerr solutions and bounds on
  the graviton mass}},
  \href{http://dx.doi.org/10.1103/PhysRevD.88.023514}{\emph{Phys. Rev.} {\bf
  D88} (2013) 023514}, [\href{http://arxiv.org/abs/1304.6725}{{\tt
  1304.6725}}].

\bibitem{Hobbs:2017oam}
G.~Hobbs and S.~Dai, \emph{{A review of pulsar timing array gravitational wave
  research}},  \href{http://arxiv.org/abs/1707.01615}{{\tt 1707.01615}}.

\bibitem{deRham:2014naa}
C.~de~Rham, L.~Heisenberg and R.~H. Ribeiro, \emph{{On couplings to matter in
  massive (bi-)gravity}},
  \href{http://dx.doi.org/10.1088/0264-9381/32/3/035022}{\emph{Class. Quant.
  Grav.} {\bf 32} (2015) 035022}, [\href{http://arxiv.org/abs/1408.1678}{{\tt
  1408.1678}}].

\bibitem{Khmelnitsky:2013lxt}
A.~Khmelnitsky and V.~Rubakov, \emph{{Pulsar timing signal from ultralight
  scalar dark matter}},
  \href{http://dx.doi.org/10.1088/1475-7516/2014/02/019}{\emph{JCAP} {\bf 1402}
  (2014) 019}, [\href{http://arxiv.org/abs/1309.5888}{{\tt 1309.5888}}].

\bibitem{Will:2014kxa}
C.~M. Will, \emph{{The Confrontation between General Relativity and
  Experiment}}, \href{http://dx.doi.org/10.12942/lrr-2014-4}{\emph{Living Rev.
  Rel.} {\bf 17} (2014) 4}, [\href{http://arxiv.org/abs/1403.7377}{{\tt
  1403.7377}}].

\bibitem{Murata:2014nra}
J.~Murata and S.~Tanaka, \emph{{A review of short-range gravity experiments in
  the LHC era}},
  \href{http://dx.doi.org/10.1088/0264-9381/32/3/033001}{\emph{Class. Quant.
  Grav.} {\bf 32} (2015) 033001}, [\href{http://arxiv.org/abs/1408.3588}{{\tt
  1408.3588}}].

\bibitem{deRham:2016nuf}
C.~de~Rham, J.~T. Deskins, A.~J. Tolley and S.-Y. Zhou, \emph{{Graviton Mass
  Bounds}}, \href{http://dx.doi.org/10.1103/RevModPhys.89.025004}{\emph{Rev.
  Mod. Phys.} {\bf 89} (2017) 025004},
  [\href{http://arxiv.org/abs/1606.08462}{{\tt 1606.08462}}].

\bibitem{Baker:2017hug}
T.~Baker, E.~Bellini, P.~G. Ferreira, M.~Lagos, J.~Noller and I.~Sawicki,
  \emph{{Strong constraints on cosmological gravity from GW170817 and GRB
  170817A}},
  \href{http://dx.doi.org/10.1103/PhysRevLett.119.251301}{\emph{Phys. Rev.
  Lett.} {\bf 119} (2017) 251301}, [\href{http://arxiv.org/abs/1710.06394}{{\tt
  1710.06394}}].

\bibitem{Merkowitz:2010kka}
S.~M. Merkowitz, \emph{{Tests of Gravity Using Lunar Laser Ranging}},
  \href{http://dx.doi.org/10.12942/lrr-2010-7}{\emph{Living Rev. Rel.} {\bf 13}
  (2010) 7}.

\bibitem{Bizon:2011gg}
P.~Bizon and A.~Rostworowski, \emph{{On weakly turbulent instability of anti-de
  Sitter space}},
  \href{http://dx.doi.org/10.1103/PhysRevLett.107.031102}{\emph{Phys. Rev.
  Lett.} {\bf 107} (2011) 031102}, [\href{http://arxiv.org/abs/1104.3702}{{\tt
  1104.3702}}].

\bibitem{Thorne:1980ru}
K.~S. Thorne, \emph{{Multipole Expansions of Gravitational Radiation}},
  \href{http://dx.doi.org/10.1103/RevModPhys.52.299}{\emph{Rev. Mod. Phys.}
  {\bf 52} (1980) 299--339}.

\bibitem{Maggiore:1900zz}
M.~Maggiore, \emph{{Gravitational Waves. Vol. 1: Theory and Experiments}}.
\newblock Oxford Master Series in Physics. Oxford University Press, 2007.

\end{thebibliography}\endgroup
\bibliographystyle{JHEP}

\end{document}